\RequirePackage[l2tabu, orthodox]{nag}
\documentclass[aps,pre,notitlepage,amsmath,amssymb,amsfonts,nobibnotes,nofootinbib,superscriptaddress]{revtex4-1}
\setcitestyle{authoryear,round}
\makeatletter
\def\p@subsection{}
\def\p@subsubsection{}
\makeatother

\usepackage[T1]{fontenc}
\usepackage[english]{babel}
\usepackage{lmodern}
\usepackage{microtype}
\usepackage{mathtools}
\usepackage[utf8]{inputenc}
\usepackage{graphicx}

\usepackage{cleveref}
\crefname{appsec}{Appendix}{Appendices}
\crefname{section}{Section}{Sections}

\usepackage{bm}
\let\vec\bm
\newcommand{\tvec}[1]{\tilde{\vec{#1}}}

\usepackage{mleftright}
\let\left\mleft
\let\right\mright

\usepackage{tikz}
\usepackage{tikz-3dplot}
\usetikzlibrary{calc}

\newcommand{\pder}[2]{\frac{\partial#1}{\partial#2}}
\newcommand{\pdder}[2]{\frac{\partial^2#1}{\partial#2^2}}
\newcommand{\der}[2]{\frac{d#1}{d#2}}

\newcommand{\sums}{\sum_s}

\newcommand{\va}{v_a}
\newcommand{\vt}{v_t}

\newcommand{\pe}{p_e}
\newcommand{\ue}{\vec{u}_e}
\newcommand{\tue}{\tvec{u}_e}

\newcommand{\omc}{\omega_c}
\newcommand{\omcs}{\omega_{cs}}
\newcommand{\vomc}{\vec{\omega}_c}

\newcommand{\ome}{\bar{\omega}_c}

\newcommand{\omg}{\omega_g}
\newcommand{\omgs}{\omega_{gs}}
\newcommand{\vomg}{\vec{\omega}_g}

\newcommand{\ex}{\vec{e}_x}
\newcommand{\ey}{\vec{e}_y}
\newcommand{\ez}{\vec{e}_z}

\newcommand{\vc}{\mathring{v}}
\newcommand{\vvc}{\vec{\mathring{v}}}

\newcommand{\cm}{\frac{e}{m}}

\newcommand{\vfreq}{\nu}

\newcommand{\qs}{e_s}

\newcommand{\rc}{r_c}

\begin{document}

\title{Linear Vlasov Theory in the Shearing Sheet Approximation \\
  with Application to the Magneto-Rotational Instability}

\author{Tobias Heinemann}
\email{tobias.heinemann@gmail.com}
\affiliation{Department of Astronomy,
  University of California, Berkeley, CA 94720, USA}
\affiliation{Kavli Institute for Theoretical Physics,
  University of California, Santa Barbara, CA 93106, USA}
\author{Eliot Quataert}
\email{eliot@berkeley.edu}
\affiliation{Department of Astronomy,
  University of California, Berkeley, CA 94720, USA}

\date{\today}

\begin{abstract}

We derive the conductivity tensor for axisymmetric perturbations of a hot,
collisionless, and charge-neutral plasma in the shearing sheet approximation.
Our results generalize the well-known linear Vlasov theory for uniform plasmas
to differentially rotating plasmas and can be used for wide range of kinetic
stability calculations. We apply these results to the linear theory of the
magneto-rotational instability (MRI) in collisionless plasmas. We show
analytically and numerically how the general kinetic theory results derived
here reduce in appropriate limits to previous results in the literature,
including the low frequency guiding center (or ``kinetic MHD'') approximation,
Hall MHD, and the gyro-viscous approximation. We revisit the cold plasma model
of the MRI and show that, contrary to previous results, an initially
unmagnetized collisionless plasma is linearly stable to axisymmetric
perturbations in the cold plasma approximation. In addition to their
application to astrophysical plasmas, our results provide a useful framework
for assessing the linear stability of differentially rotating plasmas in
laboratory experiments.

\end{abstract}

\maketitle

\section{Introduction}

The shearing sheet is a widely used model for studying the local dynamics of a
differentially rotating disk
\citep[e.g.][]{Hill1878,Goldreich1965,Hawley1996}. The shearing sheet
approximation isolates the key \emph{local} physics of rotating disks,
independent of global boundary conditions. Studies of the shearing sheet have
provided valuable insight into gravitational instability in galactic and
protostellar disks, excitation of shearing waves and spiral structure, and
angular momentum transport by turbulence driven by the magneto-rotational
instability (MRI).

In studies of self-gravitating stellar systems (in particular, galactic disks)
the linear theory of the shearing sheet has been widely developed using both
kinetic approaches as well as fluid approximations. However, despite the
astrophysical importance of low collisionality ionized accretion flows onto
compact objects \citep{Rees1982,Yuan2014}, there is no existing comprehensive
kinetic treatment of the linear theory of a collisionless \emph{plasma} in the
shearing sheet approximation. That is the aim of this paper. In particular, we
derive the general conductivity tensor for such a plasma. We then apply this
result to study the MRI in kinetic theory.

The analysis in this paper can be viewed as the generalization to a plasma of
the previous linear kinetic shearing sheet calculations for self-gravitating
stellar systems \citep[e.g.][]{Julian1966}. It is likewise a generalization to
a differentially rotating plasma of the standard linear Vlasov theory of
uniform plasmas \citep[e.g.][]{Ichimaru1973,Krall1973,Stix1992}. Moreover,
although we specialize to the MRI in the latter part of the paper, our results
are likely to be useful for understanding other aspects of differentially
rotating plasmas in kinetic theory. 

Our study of the MRI in kinetic theory generalizes and extends previous work.
In particular, \citet{Quataert2002} studied the MRI in the ``kinetic MHD'' (or
guiding center) approximation, which averages over the cyclotron motion of
particles. They also focused solely on kinetic ions, neglecting the electron
dynamics under the assumption that the electrons are much colder than the
protons. The formalism derived in this paper relaxes both of these
restrictions. Our analysis contains as various limits many non-ideal MHD
studies of the MRI, in particular those assessing the importance of the Hall
effect \citep[\cref{sec:cold-ions},][]{Wardle1999} and gyro-viscous forces
\citep[\cref{sec:gyro-viscous},][]{Ferraro2007}. We disagree, however, with
the cold plasma analysis of \citet{Krolik2006}, see \cref{sec:cold}.

This paper is organized as follows. We begin by providing a brief overview of
kinetic theory in the  shearing sheet approximation (\cref{sec:vlasov-sheet}).
We then describe in some detail the equilibrium solution of the shearing sheet
equations, emphasizing in particular the properties of the equilibrium
distribution function (\cref{sec:steady-state}).
\Cref{sec:conductivity-tensor} contains the key derivation of the conductivity
tensor and the linear dispersion relation in the shearing sheet and shows how
those can be derived from standard uniform plasma results in the literature by
an appropriate set of coordinate transformations.
\Cref{sec:cold,sec:MRI-finite-cyc} derive analytically and determine
numerically the MRI dispersion relation in various limits and discuss the
connection between our results and previous derivations in the literature.
Finally, \cref{sec:conclusions} summarizes our key results. The Appendices
\labelcref{app:unperturbed-hamiltonian} through \labelcref{app:vlasov-fluid}
contain various technical aspects of the calculations.

\section{Vlasov description of charged particle dynamics in the shearing
sheet}
\label{sec:vlasov-sheet}

The collisionless evolution of charged particles in a rotating frame of
reference is described by the Vlasov equation
\begin{equation}
  \label{eq:vlasov-sheet}
  \pder{f_s}{t} + \vec{v}\cdot\nabla f_s
  + \frac{\qs}{m_s}(\vec{E} + \vec{v}\times\vec{B})\cdot\pder{f_s}{\vec{v}}
  - (2\vec{\Omega}\times\vec{v} + \nabla\psi)\cdot\pder{f_s}{\vec{v}} = 0.
\end{equation}
Here, $f_s(\vec{r},\vec{v},t)$ is the one-particle distribution function. This
is normalized such that the number density $n_s$ and bulk velocity $\vec{u}_s$
are given by
\begin{equation}
  \label{eq:moments}
  n_s = \int d^3v\,f_s
  \quad\textrm{and}\quad
  \vec{u}_s = \int d^3v\,\vec{v}\hat{f}_s,
\end{equation}
where $\hat{f}_s=f_s/n_s$ is the reduced distribution function.

The tidal potential $\psi$ is the sum of the centrifugal potential and the
gravitational potential. In the shearing sheet approximation
\citep{Goldreich1965}, this is given by
\begin{equation}
  \label{eq:tidal-potential}
  \psi = -q\Omega^2 x^2 + \frac{1}{2}\vfreq^2 z^2.
\end{equation}
The angular velocity $\Omega$ is determined from radial force balance. We
assume that in equilibrium, the magnetic field is frozen into a bulk flow
$\vec{u}$ common to all species ($\vec{E}+\vec{u}\times\vec{B}=0$). If the
gravitational potential is due to a point mass and radial pressure support is
negligible, then $q=3/2$ (Keplerian rotation) and the vertical frequency
$\nu=\Omega$.

With the tidal potential as given in \cref{eq:tidal-potential}, the dynamics
are invariant with respect to translations along $y$. As we will verify below,
this implies conservation of canonical angular momentum. On the contrary,
there is no invariance with respect to mere translations along $x$. Instead,
such translations need to be followed by a Galilean boost along $y$, to wit
\begin{equation}
  \label{eq:shear-symmetry}
  x \mapsto x + x_0,\quad
  y \mapsto y - q\Omega t x_0,
\end{equation}
cf.~\citet{Wisdom1988}.

\section{The steady state}
\label{sec:steady-state}

In light of the symmetries discussed in the previous section, we may seek an
equilibrium in which the number densities are uniform in $x$ and $y$ and in
which the bulk motion of each plasma species is given by
\begin{equation}
  \label{eq:shear-flow}
  \vec{u} = -q\Omega x\ey.
\end{equation}
This shear flow is the local representation of the global differential
rotation. We take the magnetic field to be uniform. Ampère's law
$\mu_0\vec{J}=\nabla\times\vec{B}$ then implies that the electric current
density vanishes, which is consistent with the bulk flow
\labelcref{eq:shear-flow} being common to all species. The frozen-in condition
gives rise to an electric field $\vec{E}=q\Omega{}x\ey\times\vec{B}$. Given
this, the radial magnetic field $B_x$ must vanish according to Ferraro's
theorem \citep{Ferraro1937}. The equilibrium electromagnetic fields are thus
given by
\begin{align}
  \label{eq:equilibrium-eb-fields}
  \vec{E} = q\Omega xB_z\ex
  \quad\textrm{and}\quad
  \vec{B} = B_y\ey + B_z\ez.
\end{align}

Let us now consider the motion of an individual particle with charge-to-mass
ratio $e/m$. Given \cref{eq:equilibrium-eb-fields}, the equation of motion may
be written as
\begin{equation}
  \label{eq:particle-eom}
  \der{\vec{v}}{t} =
  \cm(\vec{E} + \vec{v}\times\vec{B})
  - (2\vec{\Omega}\times\vec{v} + \nabla\psi) =
  (\vec{v} + q\Omega x\ey)\times\vec{S} - \vfreq^2 z\ez,
\end{equation}
where we have introduced the vector
\begin{equation}
  \label{eq:spin}
  \vec{S} = \cm\vec{B} + 2\vec{\Omega}.
\end{equation}
Note that for neutral particles we recover the equation of motion for test
particles in Hill's approximation \citep{Hill1878}.

The equation of motion \labelcref{eq:particle-eom} is generated by the
Hamiltonian\footnote{Here and in the following we will refer to both
  $\mathcal{H}$ and $2\mathcal{H}$ as the Hamiltonian. The same goes for any
  energy-like quantity.}
\begin{equation}
  \label{eq:hamiltonian}
  2\mathcal{H} = p_x^2 + {(p_y - S_z x)}^2 + {(p_z + S_y x)}^2
  - q\Omega S_z x^2 + \vfreq^2 z^2,
\end{equation}
see \cref{app:unperturbed-hamiltonian} for a derivation. Expressing this in
terms of the particle velocity $\vec{v}=\partial\mathcal{H}/\partial\vec{p}$
yields
\begin{equation}
  \label{eq:energy-integral}
  2\mathcal{E} = \vec{v}^2 - q\Omega S_z x^2 + \vfreq^2 z^2,
\end{equation}
which we will refer to as the energy integral or simply the energy. According
to Jeans' theorem, any distribution function that depends only on the
integrals of motion is in equilibrium. Thus, any $f(\mathcal{E})$ is in
equilibrium. However, any such distribution function describes an equilibrium
in which the plasma is not rotationally supported but pressure-supported
against the tidal force: there is no mean flow and the density is non-uniform
in $x$. An equilibrium distribution function that is compatible with the
underlying assumptions of the shearing sheet is obtained as follows.

Inspection of \cref{eq:hamiltonian} immediately reveals that $y$ is a cyclic
coordinate. A second integral of motion (in addition to the energy integral)
is thus given by the conjugate momentum $p_y$. We will henceforth refer to
$p_y=v_y+S_z{}x$ as the (canonical) angular momentum
\citep[cf.][]{Wisdom1988}. For a given $p_y$, the dynamics may be described in
the reduced phase space $(x,z,p_x,p_z$). A particle traveling in the mid-plane
at the local shear velocity \labelcref{eq:shear-flow} corresponds to a
stationary point in this reduced phase space. To see this, we note that the
first variation of $\mathcal{H}$ is given by
\begin{equation}
  \delta\mathcal{H} = v_x\delta p_x + v_z\delta p_z
  + [S_y v_z - S_z(v_y + q\Omega x)]\delta x + \vfreq^2 z\delta z.
\end{equation}
A stationary point is reached if all coefficients in this expression vanish
identically. This is the case if and only if
\begin{equation}
  z = 0 \quad\textrm{and}\quad v_x = v_y + q\Omega x = 0.
\end{equation}
These orbits are the local representation of global circular orbits (see
\cref{app:global}). In the following this will be implied and we will refer to
them simply as circular orbits.

For a given angular momentum $p_y$, the energy of a particle on a circular
orbit is
\begin{equation}
  2\mathcal{E}_c = \frac{p_y^2}{1 - S_z/q\Omega}.
\end{equation}
Note that $\mathcal{E}_c$ depends only on the angular momentum and is thus an
integral of motion. We now define the \emph{gyration energy}
$\mathcal{K}=\mathcal{E}-\mathcal{E}_c$ as the difference between a particle's
energy and the energy of a hypothetical particle with the same angular
momentum but on a circular orbit. The gyration energy is thus
\begin{equation}
  \label{eq:gyration-energy}
  2\mathcal{K} =
  v_x^2 + \frac{{(v_y + q\Omega x)}^2}{1-\Delta} + v_z^2 + \vfreq^2 z^2,
\end{equation}
where we have introduced the \emph{tidal anisotropy}
\begin{equation}
  \label{eq:tidal-anisotropy}
  \Delta = q\Omega/S_z.
\end{equation}
The gyration energy $\mathcal{K}$ is the natural generalization of the
epicyclic energy \citep{Shu1969} to charged particles. Inspection of
\cref{eq:gyration-energy} reveals that any distribution function of the form
$f(\mathcal{K})$ describes a rotationally supported plasma with a number
density uniform in $x$ and $y$ and a bulk flow equal to \cref{eq:shear-flow}.
Note that $\mathcal{K}$ is manifestly positive definite as long as $\Delta<1$.
In this case the circular orbit thus minimizes the energy for a given angular
momentum. We discuss the case $\Delta\ge1$ in \cref{sec:delta} below. For now
we will assume that $\Delta<1$.

A third integral of motion is easily derived in two important special cases.
For a purely vertical magnetic field ($\vec{B}=B_z\ez$), the horizontal
degrees of freedom are decoupled from the vertical ones. The corresponding
integrals of motion are
\begin{equation}
  2\mathcal{K}_\perp =
  v_x^2 + \frac{{(v_y + q\Omega x)}^2}{1-\Delta}
  \quad\textrm{and}\quad
  2\mathcal{E}_z = v_z^2 + \vfreq^2 z^2.
\end{equation}
The inhomogeneity in $z$ can be dealt with using an action-angle formalism
\citep[e.g.][]{Kaufman1971,Kaufman1972}.

In the remainder of this paper we will, however, make no assumption about the
inclination of $\vec{B}$ in the $yz$-plane but instead assume that $\vfreq=0$
so that the plasma is homogeneous in $z$.\footnote{Note that there is of
  course overlap between the two cases.} Neglecting stratification is a
commonly made simplification that greatly eases the analysis because we may
assume that any disturbance of the equilibrium is periodic in $z$. In order to
integrate the particle equation of motion in this case, we introduce a new
velocity $\vvc$ whose components are given by
\begin{equation}
  \label{eq:gyrotropic-velocity}
  \vc_x = v_x,\quad
  \vc_y = \frac{v_y + q\Omega x}{\sqrt{1-\Delta}},\quad
  \vc_z = v_z.
\end{equation}
Expressed in terms of $\vvc$, the equation of
motion~\labelcref{eq:particle-eom} is
\begin{align}
  \label{eq:orbit-eom-vc}
  d\vvc/dt = \vvc\times\vomg,
\end{align}
where the \emph{gyro-frequency vector}
\begin{equation}
  \label{eq:gyro-frequency-vector}
  \vomg = S_y\ey + S_z\sqrt{1-\Delta}\,\ez.
\end{equation}
The gyration energy defined in \cref{eq:gyration-energy} is simply
\begin{equation}
  \label{eq:gyration-energy-vc}
  2\mathcal{K} = \vc^2.
\end{equation}

The salient feature of having introduced $\vvc$ is that \cref{eq:orbit-eom-vc}
takes the same form as in a uniform plasma, with the gyro-frequency vector
$\vomg$ defined in \cref{eq:gyro-frequency-vector} playing the role of the
cyclotron frequency vector $\vomc=e\vec{B}/m$. Let us therefore adopt a
cylindrical coordinate system $(\vc_\parallel,\vc_\perp,\vartheta)$ in
$\vvc$-space that is aligned with $\vomg$. In this and in the following
section, the labels $\perp$ and $\parallel$ will always refer to the direction
of $\vomg$. The steady state Vlasov equation is
\begin{equation}
  [(\vec{v} + q\Omega x\ey)\times\vec{S}]\cdot\pder{f}{\vec{v}}
  = (\vvc\times\vomg)\cdot\pder{f}{\vvc} = -\omg\pder{f}{\vartheta} = 0,
\end{equation}
where the square of the gyration frequency $\omg$ is given by
\begin{equation}
  \label{eq:gyro-frequency}
  \omg^2 = S_y^2 + S_z(S_z - q\Omega).
\end{equation}
The equilibrium distribution function is thus gyrotropic in $\vvc$-space,
i.e.\ of the form $f(\vc_\perp,\vc_\parallel)$. For this reason we will refer
to $\vvc$ as the \emph{gyrotropic velocity}.

The general form $f(\vc_\perp,\vc_\parallel)$ includes distribution functions
of the form $f(\mathcal{K})$. Such distribution functions are isotropic in
$\vvc$-space but anisotropic in $\vec{v}$-space (unless $q=0$ or
$e/m\to\infty$). This anisotropy is measured by $\Delta$ as defined in
\cref{eq:tidal-anisotropy}.\footnote{If
  $\mathbf{P}=m\int\!d^3v\,(\vec{v}-\vec{u})(\vec{v}-\vec{u})f(\mathcal{K})$
  denotes the pressure tensor, then $(P_{xx}-P_{yy})/P_{xx}=\Delta$ and
  $P_{zz}=P_{xx}$.} It always lies within the orbital plane and is inevitable
in the presence of the tidal force, hence the name \emph{tidal anisotropy}. By
contrast, the anisotropy due to different temperatures in the direction
parallel and perpendicular to $\vomg$ is a free parameter and is distinct from
the tidal anisotropy.\footnote{One might expect that the tidal anisotropy
  alone can give rise to well known plasma instabilities such as the fire hose
  instability or the mirror instability. We note, however, that the tidal
  anisotropy is unusual in the sense that it is, in general, not aligned with
  the magnetic field. Indeed, for a vertical field, the tidal anisotropy lies
  in the plane \emph{perpendicular} to the magnetic field.}

\subsection{The electromagnetic Schwarzschild distribution}

Throughout much of the subsequent analysis we will make no assumptions about
the specific form of $f(\vc_\perp,\vc_\parallel)$. Starting with section
\cref{sec:schwarzschild-response} we will, however, assume that the
distribution function is given by a Maxwellian in $\vvc$-space, to wit
\begin{equation}
  \label{em:schwarzschild}
  f(\mathcal{K}) =
  \frac{n\exp(-\mathcal{K}/\vt^2)}{{(2\pi)}^{3/2}\vt^3\sqrt{1-\Delta}},
\end{equation}
where $\vt=\mathrm{const}$ is the thermal velocity. In the guiding center
limit ($e/m\to\infty$), this reduces to a drifting Maxwellian in
$\vec{v}$-space \citep[as used in e.g.][]{Quataert2002}. Indeed, as
$e/m\to\infty$, the tidally induced anisotropy $\Delta\sim{}m/e$ disappears.
From the definitions given in
\cref{eq:gyrotropic-velocity,eq:gyro-frequency-vector} it then follows that
$\vvc\approx\vec{v}+q\Omega{}x\ey$ and $\vomg\approx\vomc$. In the limit, any
distribution function of the form $f(\vc_\perp,\vc_\parallel)$ thus becomes a
drifting gyrotropic distribution in $\vec{v}$-space, with the anisotropy being
due to the magnetic field only.

It is also instructive to consider neutral particles ($e=0$). In this case the
tidal anisotropy $\Delta=q/2$ and the gyration frequency $\omg=\kappa$, where
$\kappa=\sqrt{2(2-q)}\,\Omega$ is the epicyclic frequency. The distribution
function \labelcref{em:schwarzschild} thus reduces to the Schwarzschild
distribution \citep[see e.g.][]{Julian1966}. For this reason and for lack of a
better name, we will refer to $f(\mathcal{K})$ as given in
\cref{em:schwarzschild} as the \emph{electromagnetic Schwarzschild
distribution}.

\subsection{The case $\Delta>1$}
\label{sec:delta}

In the above derivation of the equilibrium distribution function we have
assumed that the tidal anisotropy \mbox{$\Delta=q\Omega/S_z<1$}. We have done
so because several quantities are singular if $\Delta=1$. The validity of our
analysis is also called into question when $\Delta>1$. In this case the
gyration energy is not sign definite, in which case a gyrating particle may
reach arbitrarily high velocities for any finite $\mathcal{K}$. One
consequence of this is that the electromagnetic Schwarzschild distribution
given in \cref{em:schwarzschild} is not normalizable unless a seemingly
arbitrary cutoff in velocity space is introduced.

What is the significance of $\Delta>1$? To address this question, let us first
consider neutral particles. In this case, the tidal anisotropy $\Delta=q/2$
and the condition $\Delta>1$ corresponds to
\begin{equation}
  \label{eq:rayleigh-criterion}
  \phantom{\qquad(\textrm{neutral particles})}
  \kappa^2 = 2(2-q)\Omega^2 < 0
  \qquad(\textrm{neutral particles}).
\end{equation}
We thus recover the familiar Rayleigh criterion. As is well known, circular
orbits are unstable if \cref{eq:rayleigh-criterion} is satisfied. The local
manifestation of this is that unperturbed orbits as described by
\cref{eq:particle-eom} are hyperbolic if $\kappa^2<0$. It is doubtful that the
shearing sheet approximation can be made sense of in this case. It certainly
would not describe the local dynamics of a disk.

The above discussion of neutral particle generalizes straightforwardly to
charged particles orbiting in a \emph{vertical} magnetic field, in which case
the analogue of \cref{eq:rayleigh-criterion} is
\begin{equation}
  \label{eq:rayleigh-criterion-vertical}
  \phantom{\qquad(\textrm{vertical magnetic field})}
  \omg^2 < 0
  \qquad(\textrm{vertical magnetic field}),
\end{equation}
where the gyration frequency is defined in \cref{eq:gyro-frequency}. By
analogy with the neutral case, we are led to conclude that
\cref{eq:rayleigh-criterion-vertical} is the criterion for the motion of a
\emph{charged} particle on a circular orbit to be unstable. In
\cref{app:global} we shed light on the stability of circular orbits from a
global perspective and confirm that this conclusion is indeed correct for
particles orbiting in the mid-plane of the disk.

The question of orbital stability becomes more involved if the magnetic field
has a toroidal component, in which case the magnetic field is misaligned with
the rotation axis. In this case, $\Delta>1$ corresponds to
\begin{equation}
  \label{eq:rayleigh-criterion-inclined}
  \phantom{\qquad(\textrm{inclined magnetic field})}
  \omg^2 < S_y^2
  \qquad(\textrm{inclined magnetic field}),
\end{equation}
where we remind the reader that $S_y=eB_y/m$.
\Cref{eq:rayleigh-criterion-inclined} says that if the magnetic field has a
toroidal component, then there is a region of parameter space,
$1<\Delta<1+(S_y/S_z){}^2$, where circular orbits are linearly stable even
though they do \emph{not} minimize the energy. In \cref{app:global} we
speculate that in spite of \cref{eq:rayleigh-criterion-inclined}, circular
orbits are destabilized by dissipation if the tidal anisotropy is in the range
$1<\Delta<1+(S_y/S_z){}^2$, and that $\Delta<1$ is always the relevant
criterion for orbital stability in realistic systems. In the main text we
will, however, simply ignore the subtlety associated with this corner case and
always take $\Delta<1$.

We finally note that the marginal case $\Delta=1$ corresponds to circular
orbits having constant canonical angular momentum. This statement is true
regardless of whether the magnetic field is inclined or not, see the global
analysis in \cref{app:global}. For neutral particles we recover the well known
result that circular orbits are stable if the angular momentum is a decreasing
function of radius.

\section{Linear theory}
\label{sec:linear-theory}

In this section we calculate the conductivity tensor of a collisionless plasma
in the shearing sheet approximation. The conductivity tensor is obtained from
the solution of the linearized Vlasov equation and establishes a linear
relationship between the perturbed current density and the perturbed electric
field. This together with an independent relationship between current and
electric field obtained from Maxwell's equations leads immediately to the
dispersion relation of a collisionless plasma.

The calculation of the conductivity tensor of a uniform collisionless plasma
is the subject of many textbooks
\citep[e.g.][]{Krall1973,Ichimaru1973,Stix1992}. In generalizing this
calculation to the shearing sheet, our strategy will be to manipulate the
governing equations in a way that allows us to leverage as many results as
possible from the (already tedious) uniform plasma calculation.

Let us first write the dynamical equations, i.e.\ the Vlasov equation
\labelcref{eq:vlasov-sheet} and Faraday's law
$\partial\vec{B}/\partial{}t+\nabla\times\vec{E}=0$, in a form that is
manifestly invariant with respect to the Galilean transformation
\labelcref{eq:shear-symmetry}. For this we introduce the \emph{relative}
velocity and electric field
\begin{equation}
  \label{eq:relative-quantities}
  \tvec{v} = \vec{v} + q\Omega x\ey
  \quad\textrm{and}\quad
  \tvec{E} = \vec{E} - q\Omega x\ey\times\vec{B}.
\end{equation}
Note that $\tvec{v}$ and $\tvec{E}$ are the velocity and electric field as
seen by an observer that is locally at rest with respect to the background
shear flow. Expressed in terms of the relative velocity and electric field,
the Vlasov equation is given by
\begin{equation}
  \label{eq:vlasov-relative}
  \left(\pder{}{t} - q\Omega x\pder{}{y}\right) f_s
  + \tilde{\vec{v}}\cdot\nabla f_s
  + \frac{\qs}{m_s}(\tilde{\vec{E}} + \tilde{\vec{v}}\times\vec{B})
  \cdot\pder{f_s}{\tilde{\vec{v}}}
  - (2\vec{\Omega}\times\tvec{v} - q\Omega\tilde{v}_x\ey + \vfreq^2 z\ez)
  \cdot\pder{f_s}{\tilde{\vec{v}}} = 0,
\end{equation}
where it is understood that $\partial/\partial{}x$ is now to be evaluated at
fixed $\tilde{v}_y$. Faraday's law expressed in terms of $\tvec{E}$ may be
written as
\begin{equation}
  \label{eq:faraday-relative}
  \left(\pder{}{t} - q\Omega x\pder{}{y}\right)
  \vec{B} + q\Omega B_x\ey + \nabla\times\tilde{\vec{E}} = 0.
\end{equation}
This form of Faraday's law makes explicit that the magnetic field is stretched
and advected by the background shear.

We note that \cref{eq:vlasov-relative,eq:faraday-relative} are exact. No
approximations have been made so far. We also note that the only explicit
$x$-dependence in \cref{eq:vlasov-relative,eq:faraday-relative} is multiplying
$\partial/\partial{}y$. Provided the dynamics are axisymmetric, we may thus
take the distribution function $f_s$, the magnetic field $\vec{B}$, and the
\emph{relative} electric field $\tvec{E}$ to be periodic in $x$.\footnote{More
  generally, if $\partial/\partial{}y\ne0$, then each field appearing in
  \cref{eq:vlasov-relative,eq:faraday-relative} may be written as a sum of
  so-called shearing waves \citep{Thomson1887,Goldreich1965}. In the context
  of computer simulations, the relative electric field $\tvec{E}$ as well as
  $\vec{B}$ and $f_s$ may be taken to be shearing periodic
  \citep{Lees1972,Hawley1995}.}

\subsection{The dispersion relation}
\label{sec:conductivity-tensor}

The conductivity tensor is obtained from the solution of the linearized Vlasov
equation
\begin{equation}
  \label{eq:linear-vlasov-sheet}
  \frac{D}{Dt}\delta f_s
  + \frac{\qs}{m_s}(\delta\tvec{E} + \tvec{v}\times\delta\vec{B})
  \cdot\pder{f_s}{\tvec{v}} = 0,
\end{equation}
where $D/Dt$ denotes the time derivative along unperturbed orbits discussed in
\cref{sec:steady-state}. The formal solution is given by
\begin{equation}
  \label{eq:formal-solution}
  \delta f_s(\vec{r},\tvec{v},t) = -\frac{\qs}{m_s}\int_{-\infty}^t\!dt'\,
  [\delta\tvec{E}(\vec{r}',t') + \tvec{v}'\times\delta\vec{B}(\vec{r}',t')]
  \cdot\pder{f_s}{\tvec{v}'}.
\end{equation}
This is a so-called orbit integral: $\vec{r}'(t')$ and $\tvec{v}'(t')$
describe the phase space trajectory of particle on an unperturbed orbit
passing through $(\vec{r},\tvec{v})$ at time $t$.

From now on we will assume no stratification ($\vfreq=0$) and axisymmetric
disturbances ($\partial/\partial{}y=0$). The dynamical equations then have
constant coefficients and we may assume that all disturbances depend on
position $\vec{r}$ and time $t$ as $\exp(ik_x{}x+ik_z{}z-i\omega{}t)$. For
such disturbances, Faraday's law \labelcref{eq:faraday-relative} is easily
solved for $\delta\vec{B}$, to wit
\begin{equation}
  \label{eq:linearized-faraday}
  \delta\vec{B} = \frac{\vec{k}}{\omega}\times\left[
    \left(\mathbf{1} - \frac{q\Omega}{i\omega}\ex\ey\right)
    \cdot\delta\tilde{\vec{E}}
  \right].
\end{equation}
Here, the second term in parentheses arises from the stretching term in
\cref{eq:faraday-relative}. Substituting \cref{eq:linearized-faraday} into
\cref{eq:formal-solution}, taking the first moment, multiplying by $\qs{}n_s$,
and summing over species yields the perturbed current density
\begin{equation}
  \label{eq:perturbed-current}
  \delta\vec{J}(\vec{r},t)
  = -\frac{1}{i\omega}\sums\frac{\qs^2n_s}{m_s}
  \int d^3v\,\int_{-\infty}^t dt'
  \tvec{v}\pder{\hat{f}_s}{\tvec{v}'}\cdot\big[
  i(\omega - \vec{k}\cdot\tvec{v}')\mathbf{1}
  + i\vec{k}\tvec{v}' + q\Omega\ex\ey
  \big]\cdot\left(\mathbf{1} - \frac{q\Omega}{i\omega}\ex\ey\right)
  \cdot\delta\tvec{E}(\vec{r}',t'),
\end{equation}
where $\hat{f}_s=f_s/n_s$ is the reduced distribution function.

\Cref{eq:perturbed-current} establishes a linear relationship between
$\delta\vec{J}$ and $\delta\tvec{E}$. For this relation to be useful we need
to carry out the orbit integral. Inspection of the unperturbed equations of
motion might give the impression that in the shearing sheet approximation,
this is a more complicated task than in a uniform plasma. However, we have
seen in the previous section that the equilibrium orbits closely resemble
those in a uniform plasma when expressed in terms of the gyrotropic velocity
$\vvc$ defined in \cref{eq:gyrotropic-velocity}. This suggests that we try to
express \cref{eq:perturbed-current} solely in terms of $\vvc$. In
\cref{app:conductivity-tensor} we show that this yields
\begin{equation}
  \label{eq:conductive-relation}
  \delta\vec{J} =
  - \frac{1}{i\omega}\sums\frac{\qs^2 n_s}{m_s}
  \left(
    \mathbf{Q}_s\cdot\mathbf{\Lambda}_s\cdot\mathbf{Q}_s + \Delta_s\ey\ey
  \right)\cdot\left(\mathbf{1} - \frac{q\Omega}{i\omega}\ex\ey\right)
  \cdot\delta\tvec{E},
\end{equation}
where the \emph{response tensor}
\begin{equation}
  \label{eq:response-tensor}
  \mathbf{\Lambda}_s =
  \int d^3v\int_0^\infty d\tau\,
  \vvc\pder{\hat{f}_s}{\vvc'}\cdot
  [i(\omega - \vec{k}\cdot\vvc')\mathbf{1} + i\vec{k}\vvc']
  \exp\{i\vec{k}\cdot(\vec{r}'-\vec{r}) + i\omega\tau\}
\end{equation}
with $\tau=t-t'$. Here, $\Delta_s=q\Omega/(\omcs{}b_z+2\Omega)$ is the tidal
anisotropy, with both $\omcs=e_s{}B/m_s$ and $b_z$ being signed quantities.
The \emph{anisotropy tensor} $\mathbf{Q}_s$ in \cref{eq:conductive-relation}
is given by
\begin{equation}
  \label{eq:anisotropy-tensor}
  \mathbf{Q}_s = \ex\ex + \ey\ey\sqrt{1-\Delta_s} + \ez\ez.
\end{equation}
While everything to the left of $\delta\tvec{E}$ on the right hand side of
\cref{eq:conductive-relation} could rightfully be referred to as the shearing
sheet conductivity tensor, we choose to define this as
\begin{equation}
  \label{eq:conductivity}
  \vec{\sigma} = -\frac{1}{i\omega}\sums\frac{\qs^2n_s}{m_s}
  \mathbf{Q}_s\cdot\mathbf{\Lambda}_s\cdot\mathbf{Q}_s.
\end{equation}
Defined in this way, it most closely resembles the uniform plasma conductivity
tensor \citep[see e.g.][]{Ichimaru1973}.

We are now in a position to write down the general dispersion relation of a
charge-neutral collisionless plasma in the shearing sheet approximation.
Inserting the solution of Faraday's law given in \cref{eq:linearized-faraday}
into Ampère's law $\mu_0\delta\vec{J}=i\vec{k}\times\delta\vec{B}$ (neglecting
the displacement current), and eliminating the perturbed current using
\cref{eq:conductive-relation} yields
\begin{equation}
  \label{eq:dispersion-relation}
  \mathbf{D}\cdot\left(
    \mathbf{1} - \frac{q\Omega}{i\omega}\ex\ey
  \right)\cdot\delta\tvec{E} = 0,
\end{equation}
where the dispersion tensor
\begin{equation}
  \label{eq:dispersion-tensor}
  \mathbf{D} =
  (k^2\mathbf{1} - \vec{k}\vec{k} - i\omega\mu_0\vec{\sigma})\va^2
  - 2q\Omega^2\sums\frac{n_s m_s}{\rho}
  \frac{\omcs/b_z}{\omcs b_z + 2\Omega}\ey\ey.
\end{equation}
Here, $\rho=\sums{}n_s{}m_s$ is the mass density, $\va^2=B^2/\mu_0\rho$ is
the square of the Alfvén speed, and we have used the charge-neutrality
condition to rewrite the last term on the right hand side.

\subsection{The plasma response}
\label{sec:plasma-response}

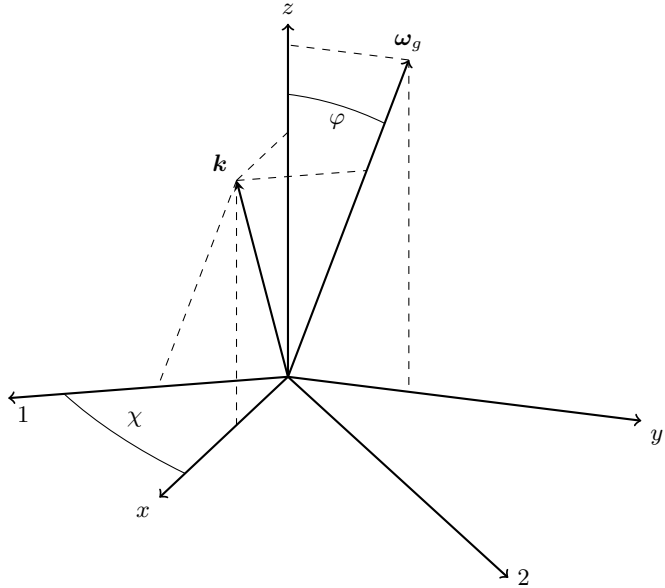
\begin{figure}
  \centering
  \tdplotsetmaincoords{70}{110}

\begin{tikzpicture}[scale=5,tdplot_main_coords]

\coordinate (O) at (0,0,0);

\draw[thick,->] (0,0,0) -- (1,0,0) node[anchor=north east]{$x$};
\draw[thick,->] (0,0,0) -- (0,1,0) node[anchor=north west]{$y$};
\draw[thick,->] (0,0,0) -- (0,0,1) node[anchor=south]{$z$};

\pgfmathsetmacro{\myphi}{20}
\pgfmathsetmacro{\klen}{0.8}
\pgfmathsetmacro{\kinc}{30}

\pgfmathsetmacro{\kz}{cos(\kinc)*\klen}
\pgfmathsetmacro{\kx}{sin(\kinc)*\klen}

\pgfmathsetmacro{\myalpha}{-atan(\kz*sin(\myphi)/\kx)}

\tdplotsetcoord{K}{\klen}{\kinc}{0}
\draw[black,dashed] (K) -- (Kyz);
\draw[black,dashed] (K) -- (Kxy);

\tdplotsetthetaplanecoords{90}
\tdplotdrawarc[tdplot_rotated_coords]{(0,0,0)}{0.8}{0}{\myphi}{anchor=north}{$\varphi$}

\tdplotsetcoord{G}{1}{\myphi}{90}
\draw[black,dashed] (G) -- (Gxz);
\draw[black,dashed] (G) -- (Gxy);

\tdplotsetrotatedcoords{90}{\myphi}{\myalpha-90}
\tdplotdrawarc[tdplot_rotated_coords]{(0,0,0)}{0.8}{0}{-\myalpha}{anchor=south
west}{$\chi$}

\tdplottransformmainrot{\kx}{0}{\kz}

\draw[thick,->,tdplot_rotated_coords] (0,0,0) -- (1,0,0)
node[anchor=north west,black]{$1$};
\draw[thick,->,tdplot_rotated_coords] (0,0,0) -- (0,1,0)
node[anchor=west,black]{$2$};
\draw[thick,->,tdplot_rotated_coords] (0,0,0) -- (0,0,1)
node[anchor=south,black]{$\bm{\omega}_g$};

\draw[tdplot_rotated_coords,dashed]
(\tdplotresx,0,\tdplotresz) -- (\tdplotresx,0,0);
\draw[tdplot_rotated_coords,dashed]
(\tdplotresx,0,\tdplotresz) -- (0,0,\tdplotresz);

\draw[-stealth,thick] (O) -- (\kx,0,\kz)
node[anchor=south east,black]{$\bm{k}$};


\end{tikzpicture}
  \caption{Illustration of the coordinate transformations
    \labelcref{eq:e3,eq:e12} used in calculating the plasma response tensor
    \labelcref{eq:response-tensor-gyrotropic}. The $(x,y,z)$-coordinate frame
    is the standard shearing sheet coordinate frame. The gyro-frequency vector
    $\vomg$ defines the $3$-axis of the $(1,2,3)$-coordinate frame that we
    refer to as the Stix frame. The $1$-axis is rotated with respect to the
    $x$-axis by an angle $\chi$ about the $3$-axis, which in turn is
    inclined with respect to the $z$-axis by an angle $\varphi$ about the
    $x$-axis. The two Euler angles $\varphi$ and $\chi$ are related through
    $k_x\tan\chi=k_z\sin\varphi$.}\label{fig:stix}
\end{figure}

From now on we will suppress the species index when there is no risk of
confusion. As promised earlier, the orbit integral in
\cref{eq:response-tensor} only involves the gyrotropic velocity. Let us adopt
a right-handed Cartesian coordinate system $(\vec{e}_1,\vec{e}_2,\vec{e}_3)$
with $\vec{e}_3=\vomg/\omg$. The unperturbed orbit in $\vvc$-space is then
simply
\begin{equation}
  \label{eq:vcirc-tau}
  \vvc' = \vc_\perp[\vec{e}_1\cos(\vartheta + \omg\tau)
  + \vec{e}_2\sin(\vartheta + \omg\tau)] + \vc_\parallel\vec{e}_3.
\end{equation}
The gyro-frequency has no component along the $x$-direction. Thus we may write
\begin{equation}
  \label{eq:e3}
  \vec{e}_3 = \ez\cos\varphi - \ey\sin\varphi.
\end{equation}
The orientation of the coordinate axes in the plane perpendicular to $\vomg$
is arbitrary at this point. We can exploit this freedom by choosing a frame in
which the wave vector $\vec{k}=k_x\ex+k_z\ez$ lies in the plane spanned by
$\vec{e}_1$ and $\vec{e}_3$. The desired orientation of the basis vectors is
achieved if $\vec{e}_1$ and $\vec{e}_2$ are defined through
\begin{align}
  \label{eq:e12}
  k_\perp\vec{e}_1 = k_x\ex + k_z\sin\varphi\,\vec{e}_3\times\ex
\end{align}
and $\vec{e}_2=\vec{e}_3\times\vec{e}_1$. Indeed, projecting the wave vector
onto the new basis shows that $\vec{k}=k_\perp\vec{e}_1+k_\parallel\vec{e}_3$,
where the perpendicular and parallel wave number are given by
\begin{equation}
  k_\perp^2 = k_x^2 + k_z^2\sin^2\varphi
  \quad\textrm{and}\quad
  k_\parallel = k_z\cos\varphi,
\end{equation}
respectively. In the following we will refer to the coordinate basis defined
by \cref{eq:e3,eq:e12} and illustrated in \cref{fig:stix} as the \emph{Stix
basis}.

In order to determine the particle orbit in position space, we note that for
axisymmetric disturbances it follows from \cref{eq:gyrotropic-velocity} that
$\vec{k}\cdot{}d\vec{r}'/dt'=\vec{k}\cdot\vec{v}'=\vec{k}\cdot\vvc'$. Working
in the Stix basis, it is easy to see that integrating this equation subject to
the condition $\vec{r}'(t)=\vec{r}$ yields
\begin{equation}
  \label{eq:k-dot-r}
  \vec{k}\cdot(\vec{r}-\vec{r}')
  = \frac{k_\perp\vc_\perp}{\omg}
  [\sin(\vartheta + \omg\tau) - \sin\vartheta]
  + k_\parallel\vc_\parallel\tau.
\end{equation}
With this we can now carry out the integrals over $\tau$ and the gyro-phase
$\vartheta$ in \cref{eq:response-tensor}. This calculation is entirely
analogous to the corresponding calculation in a uniform plasma \citep[see
e.g.][]{Ichimaru1973} modulo the substitutions
$(v_\perp,v_\parallel,\omcs)\to(\vc_\perp,\vc_\parallel,\omgs)$. The result
may be written as\footnote{The integral over velocity space in
  \cref{eq:response-tensor-gyrotropic} is understood to be
  $\int{}d^3v=2\pi\sqrt{1-\Delta}
  \int_0^\infty{}d\vc_\perp\vc_\perp\int_{-\infty}^\infty{}d\vc_\parallel$.}
\begin{equation}
  \label{eq:response-tensor-gyrotropic}
  \mathbf{\Lambda} =
  \mathbf{1} + \sum_{n=-\infty}^\infty\int d^3v\left(
    \frac{n\omg}{\vc_\perp}\pder{\hat{f}}{\vc_\perp}
    + k_\parallel\pder{\hat{f}}{\vc_\parallel}
  \right)
  \frac{\mathbf{\Pi}_n(\vc_\perp,\vc_\parallel)}
  {n\omg + k_\parallel\vc_\parallel - \omega}.
\end{equation}
In the Stix basis ($\vec{e}_1,\vec{e}_2,\vec{e}_3$), the tensors
$\mathbf{\Pi}_n$ have the matrix representation
\begingroup
\renewcommand*{\arraystretch}{2}
\begin{equation}
  \label{eq:Pi-tensor}
  \mathbf{\Pi}_n(\vc_\perp,\vc_\parallel) \doteq \left[\begin{matrix}
      \dfrac{n^2\omg^2}{k_\perp^2} J_n^2 &
      i\vc_\perp\dfrac{n\omg}{k_\perp} J^{}_n J_n' &
      \vc_\parallel\dfrac{n\omg}{k_\perp} J_n^2 \\
      -i\vc_\perp\dfrac{n\omg}{k_\perp} J^{}_n J_n' &
      \vc_\perp^2 J_n'^2 &
      -i\vc_\parallel\vc_\perp J^{}_n J_n' \\
      \vc_\parallel\dfrac{n\omg}{k_\perp} J_n^2 &
      i\vc_\parallel\vc_\perp J^{}_n J_n' &
      \vc_\parallel^2 J_n^2
  \end{matrix}\right],
\end{equation}
\endgroup
where $J_n$ is the Bessel function of order $n$. Its argument is
$k_\perp\vc_\perp/\omg$ and $J_n'$ denotes the derivative with respect to said
argument.

We stress that all we have assumed in deriving
\cref{eq:response-tensor-gyrotropic,eq:Pi-tensor} is $k_y=0$. In order to
carry out the remaining integrations over $\vc_\perp$ and $\vc_\parallel$ in
\cref{eq:response-tensor-gyrotropic}, we need to specify the particular form
of the distribution function $f(\vc_\perp,\vc_\parallel)$. In the next section
we will carry out this calculation for the electromagnetic Schwarzschild
distribution defined in \cref{em:schwarzschild}.

\subsection{Response tensor for the electromagnetic Schwarzschild distribution}
\label{sec:schwarzschild-response}

The result given in \cref{eq:response-tensor-gyrotropic} holds for general
equilibrium distribution functions of the form $f(\vc_\perp,\vc_\parallel)$.
We now specialize to the electromagnetic Schwarzschild distribution defined in
\cref{em:schwarzschild} and given by
\begin{equation}
  \label{em:schwarzschild-2}
  f(\mathcal{K}) =
  \frac{n\exp(-\mathcal{K}/\vt^2)}{{(2\pi)}^{3/2}\vt^3\sqrt{1-\Delta}},
\end{equation}
where $\vt=\mathrm{const}$ is the thermal velocity. We remind the reader that
the tidal anisotropy \mbox{$\Delta=q\Omega/(\omc{}b_z+2\Omega)$} and the
gyration energy
\begin{equation}
  \mathcal{K} = \frac{1}{2}
  \left[v_x^2 + \frac{{(v_y + q\Omega x)}^2}{1 - \Delta} + v_z^2\right] =
  \frac{\vc^2}{2}.
\end{equation}
The second equality shows that we are simply considering a plasma that is
Maxwellian in $\vvc$-space.

Given the equilibrium distribution function in \cref{em:schwarzschild-2}, the
integrals over $\vc_\perp$ and $\vc_\parallel$ that appear in the response
tensor \labelcref{eq:response-tensor-gyrotropic} can now be performed. This
calculation is again entirely analogous to the corresponding calculation in a
uniform plasma \citep[see e.g.][]{Ichimaru1973}. We may thus spare the reader
the details and simply state the result
\begin{equation}
  \label{eq:response-tensor-schwarzschild}
  \mathbf{\Lambda} =
  \sum_{n=-\infty}^\infty\frac{\zeta_0}{\zeta_n}
  \{1 - W(\zeta_n)\}\mathbf{T}_n - \zeta_0^2\vec{e}_3\vec{e}_3,
\end{equation}
where
\begin{equation}
  \label{eq:zeta-def}
  \zeta_n = \frac{\omega - n\omg}{|k_\parallel|\vt}.
\end{equation}
The $W$-function used here and in \citet{Ichimaru1973} is related to the
standard plasma dispersion function $Z$ defined by \citet{Fried1961} through
$W(\zeta)=1+\xi{}Z(\xi)$ with $\zeta=\sqrt{2}\,\xi$. In the Stix basis, the
tensors $\mathbf{T}_n$ in \cref{eq:response-tensor-schwarzschild} have the
matrix representation
\begingroup
\renewcommand*{\arraystretch}{2}
\begin{equation}
  \label{eq:T-tensor}
  \mathbf{T}_n \doteq \left[\begin{matrix}
      \dfrac{n^2}{\lambda}\Gamma_n &
      in\Gamma_n' &
      \dfrac{k_\parallel}{|k_\parallel|}
      \dfrac{n}{\sqrt{\lambda}}\,\zeta_n\Gamma_n \\
      -in\Gamma_n' &
      \dfrac{n^2}{\lambda}\Gamma_n - 2\lambda\Gamma_n' &
      -i\dfrac{k_\parallel}{|k_\parallel|}
      \sqrt{\lambda}\,\zeta_n\Gamma_n' \\
      \dfrac{k_\parallel}{|k_\parallel|}
      \dfrac{n}{\sqrt{\lambda}}\,\zeta_n\Gamma_n &
      i\dfrac{k_\parallel}{|k_\parallel|}
      \sqrt{\lambda}\,\zeta_n\Gamma_n' &
      \zeta_n^2\Gamma_n
  \end{matrix}\right].
\end{equation}
\endgroup
Here, the $\Gamma_n$ are related to the modified Bessel functions of the first
kind $I_n$ through
\begin{equation}
  \label{eq:Gamma-n}
  \Gamma_n(\lambda) = I_n(\lambda)\exp(-\lambda).
\end{equation}
Their argument in \cref{eq:T-tensor} is
\begin{equation}
  \label{eq:chi-def}
  \lambda = \frac{k_\perp^2\vt^2}{\omg^2}.
\end{equation}

\subsection{Cold plasma limit}
\label{sec:cold}

The cold plasma conductivity tensor may be obtained from the results of the
previous section by letting $\vt\to0$. In this limit we may drop all
$W$-functions in \cref{eq:response-tensor-schwarzschild}, since
$\zeta_n\to\infty$, and most of the $\Gamma_n$ in \cref{eq:T-tensor}, since
$\lambda\to0$. We are then left with the response tensor
\begin{equation}
  \label{eq:cold-response}
  \mathbf{\Lambda}_s = \vec{e}_3\vec{e}_3
  + \frac{1}{2}\sum_\pm\frac{\omega}{\omega\pm\omgs}
  (\vec{e}_1\pm i\vec{e}_2)(\vec{e}_1\mp i\vec{e}_2).
\end{equation}
The response tensor is singular for $\omega^2=\omgs^2$. This can be understood
as follows. The linearized cold plasma momentum equation in the presence of
Coriolis and tidal forces is
\begin{equation}
  \label{eq:momentum}
  \left(
    \pder{}{t} - q\Omega x\pder{}{y}
    + 2\vec{\Omega}\times\mathbf{1} - q\Omega\ey\ex
  \right)\cdot\delta\tvec{u}_s =
  \frac{\qs}{m_s}(\delta\tvec{E} + \delta\tvec{u}_s\times\vec{B}).
\end{equation}
Multiplying this with $n_s{}m_s$, summing over species, assuming axisymmetric
perturbations, and using charge-neutrality yields\footnote{Multiplying the
  right hand side of \cref{eq:momentum} by $n_s{}m_s$ and summing over species
  yields of course the linearized Lorentz force $\delta\vec{J}\times\vec{B}$.
  This vanishes for $\vec{k}=0$ by Ampère's law, in which case the solubility
  condition reduces to $\omega^2=\kappa^2$, independent of any electromagnetic
  effects. This also follows in complete generality (assuming only
  charge-neutrality) from the Vlasov equation \labelcref{eq:vlasov-relative}.}
\begin{equation}
  \label{eq:sum-cold-momentum}
  \sums\left[
    -i\omega\mathbf{1}
    + \left(2\vec{\Omega} + \frac{\qs}{m_s}\vec{B}\right)\times\mathbf{1}
    - q\Omega\ey\ex
  \right]\cdot n_s m_s\delta\tvec{u}_s = 0.
\end{equation}
This equation has a non-trivial solution (with at least one
$\delta\tvec{u}_s\ne0$) if the tensor in square brackets is singular, i.e.\ if
\begin{equation}
  i\omega(\omega^2 - \omgs^2) = 0,
\end{equation}
where the square of the gyration frequency $\omgs$ is defined in
\cref{eq:gyro-frequency}.

The response tensor in \cref{eq:cold-response} is derived by expressing the
perturbed current in terms of the perturbed electric field, which requires
inverting  the tensor in square brackets in \cref{eq:sum-cold-momentum}. The
apparent singularity in the plasma response at $\omega^2=\omgs^2$ is a
consequence of this inversion being possible only if said tensor is
non-singular, i.e.\ if $\omega^2\ne\omgs^2$. In what follows we will thus
assume that $\omega^2\ne\omgs^2$. 

Combining \cref{eq:conductivity,eq:dispersion-tensor,eq:cold-response} leads
to the dispersion relation of a cold multicomponent plasma in the shearing
sheet approximation. The dynamics of such a plasma were studied previously by
\citet{Krolik2006}. We disagree with their analysis for the following reasons.
First, they have an incorrect expression for the gyration
frequency, which can be read off from the denominator of their eq.~(3), which
disagrees with our \cref{eq:gyro-frequency}. Second, their dispersion relation
does not reproduce the Hall MRI dispersion relation
\citep{Wardle1999,Balbus2001} in the limit $m_e\to0$, as it should (and as our
dispersion relation does, see \cref{sec:cold-ions} below). Finally,
\citet{Krolik2006} claim that a differentially rotating, cold, and initially
\emph{unmagnetized} plasma is linearly unstable with a growth rate comparable
to the rotation frequency. We will now show that this is in fact not the case.

\subsubsection{Zero magnetization}
\label{sec:zero-mag}

Here we consider the stability of an initially unmagnetized cold plasma in the
shearing sheet. The dispersion relation for this case is obtained from the
general expression \cref{eq:dispersion-tensor} together with
\cref{eq:conductivity} by substituting \cref{eq:cold-response} and then
letting the magnetic field strength tend to zero. The dispersion tensor in
this limit reduces to
\begin{equation}
  \label{eq:unmagnetized-dispersion-tensor}
  \va^{-2}\ell^2\mathbf{D} =
  (k^2\mathbf{1} - \vec{k}\vec{k} - i\omega\mu_0\vec{\sigma})\ell^2
  + q/2\,\ey\ey,
\end{equation}
where $\ell$ is the mean inertial length, defined through
\begin{equation}
  \ell^{-2} = \mu_0\sums\frac{\qs^2n_s}{m_s}.
\end{equation}
The limiting form of the cold plasma conductivity tensor is
\begin{equation}
  \label{eq:zero-mag-conductivity}
  -i\omega\mu_0\vec{\sigma} =
  \ell^{-2}\mathbf{Q}\cdot\mathbf{\Lambda}\cdot\mathbf{Q},
\end{equation}
where the anisotropy tensor is given in \cref{eq:anisotropy-tensor} with
$\Delta_s=q/2$ and the cold plasma response tensor for $B=0$ is
\begin{equation}
  \label{eq:zero-mag-response}
  \mathbf{\Lambda} = \ez\ez
  + \frac{1}{2}\sum_{\pm}\frac{\omega}{\omega\pm\kappa}
  (\ex\pm i\ey)(\ex\mp i\ey).
\end{equation}
With this, setting the determinant of \cref{eq:unmagnetized-dispersion-tensor}
equal to zero yields the dispersion relation
\begin{equation}
  \frac{\omega^2{(1 + k^2\ell^2)}^2 - \kappa^2 k_z^2\ell^2(q/2 + k^2\ell^2)}
  {\omega^2 - \kappa^2} = 0.
\end{equation}
Assuming $\omega^2\ne\kappa^2$, the solution is
\begin{equation}
  \omega^2 =
  \frac{\kappa^2k_z^2\ell^2(q/2 + k^2\ell^2)}{{(1 + k^2\ell^2)}^2}.
\end{equation}
This is manifestly positive as long as $q>0$. Thus there is no instability as
long as the orbital frequency decreases radially outward.

\section{The kinetic MRI with finite ion cyclotron frequency}
\label{sec:MRI-finite-cyc}

\subsection{Cold and massless electrons}
\label{sec:cold-e}

In most plasmas, electrons are much lighter than ions, and so the mass ratio
$m_e/m_i\ll1$. If we assume that the electrons are also much colder than the
ions ($T_e/T_i\ll1$), then the electrons can be modeled as a cold massless
fluid. Expanding the cold electron conductivity tensor in the mass ratio
yields
\begin{equation}
  \label{eq:cold-electron-conductivity}
  \vec{\sigma}_e \approx -\frac{1}{i\omega}\frac{e^2n_e}{m_e}\vec{b}\vec{b}
  + \frac{en_e}{B}\vec{b}\times\mathbf{1},
\end{equation}
where we have dropped terms of order $m_e/m_i$ and higher. The first term
evidently diverges as the mass ratio goes to zero. Because this term is
proportional to $\vec{b}\vec{b}$, the dispersion tensor becomes
block-diagonal.\footnote{We note that since
  $\mathbf{D}=-i\omega\mu_0\vec{\sigma}_e\va^2+\ldots$, it follows from the
  dispersion relation given in \cref{eq:dispersion-relation} that the parallel
  component of the (relative) electric field must be at least first order in
  $m_e/m_i$. The parallel current density would otherwise diverge, a result
  that is consistent with guiding center theory \citep[e.g.][]{Grad1961}.} In
the cold and massless electron limit, the dispersion relation thus factorizes
into $\delta{}\tilde{E}_\parallel=0$ and $\det\mathbf{D}_\perp=0$, where the
perpendicular dispersion tensor
\begin{equation}
  \label{eq:perpendicular-dispersion}
  \mathbf{D}_\perp =
  {(k^2\mathbf{1} - \vec{k}\vec{k} - i\omega\mu_0\vec{\sigma}_i)}_\perp\va^2
  - i\omega\ome\vec{b}\times\mathbf{1}
  - 2q\Omega^2\sums\frac{n_s{}m_s}{\rho}\frac{\omcs b_z}{\omcs b_z + 2\Omega}
  (\vec{b}\times\ex)(\vec{b}\times\ex).
\end{equation}
We note that throughout this section, the labels $\perp$ and $\parallel$ refer
to the direction of $\vec{b}$, \emph{not} to the gyro-frequency vector $\vomg$
as in \cref{sec:steady-state,sec:linear-theory}. In
\cref{eq:perpendicular-dispersion}, $\vec{\sigma}_i$ is the (multi-component)
ion conductivity tensor and
\begin{equation}
  \ome = \frac{en_e B}{\rho} = \sums\frac{n_s m_s}{\rho}\omcs
\end{equation}
is the (mass density weighted) mean ion cyclotron frequency.

In \cref{app:vlasov-fluid} we derive the dispersion relation of a so-called
Vlasov-fluid in the shearing sheet. Vlasov-fluid theory \citep{Freidberg1972}
assumes that the electrons are massless and isotropic, but not necessarily
cold. The dispersion relation derived above agrees with the Vlasov-fluid
dispersion relation for $T_e=0$, see \cref{eq:vlasov-fluid-dispersion-tensor}.

In the remainder of this section we will assume that there is only one ion
species. The perpendicular dispersion tensor in this case is given by
\begin{equation}
  \label{eq:perpendicular-dispersion-single}
  \mathbf{D}_\perp =
  \va^2{(k^2\mathbf{1} - \vec{k}\vec{k})}_\perp
  + \omc^2{(\mathbf{Q}\cdot\mathbf{\Lambda}\cdot\mathbf{Q})}_\perp
  - i\omega\omc\vec{b}\times\mathbf{1}
  - 2q\Omega^2\omc b_z/S_z
  (\vec{b}\times\ex)(\vec{b}\times\ex).
\end{equation}
We focus on the Vlasov-fluid limit in this section to make contact with the
existing kinetic theory calculations of the MRI, all of which have assumed
cold electrons. We stress, however, that the results derived in the previous
section are fully general and can be used to quantify the effects of kinetic
electrons on the MRI\@. The cold electron approximation used here and in
previous studies formally requires that $\zeta_n\rightarrow\infty$ for all $n$
(\cref{sec:cold}). However,
$\zeta_0=\omega/|k_\parallel|v_{te}\sim{}\va/v_{te}$ for the MRI (taking
$\omega\sim{}k_\parallel\va$). Thus $\zeta_0\gg{}1$ requires
$T_e/T_i\ll(m_e/m_i)\beta_i^{-1}$. This is in general \emph{not} satisfied so
that the cold electron approximation is not formally applicable when studying
the MRI in hot accretion flows. In future work, we will study the effect of
(hot) kinetic electrons on the MRI in collisionless plasmas.

\subsection{The guiding center limit}
\label{sec:guiding-center}

In the limit $e/m\to\infty$ we should be able to recover the results of
\citet{Quataert2002} who analyzed the stability of a differentially rotating
plasma in the shearing sheet approximation using the so-called kinetic MHD
equations \citep[see e.g.][]{Grad1961,Kulsrud1983,Hazeltine2004}. We start by
expanding the response tensor in powers of $m/e$. Keeping mostly leading order
terms, this yields
\begin{equation}
  \mathbf{\Lambda} \approx
  -\frac{\omega^2}{\omc^2}(\mathbf{1} - \vec{e}_3\vec{e}_3)
  + \frac{i\omega}{\omc}\left[1 - \frac{(4-q)\Omega b_z}{2\omc}\right]
  \vec{e}_3\times\mathbf{1}
  - \frac{2(W-1)k_\perp^2\vt^2}{\omc^2}\vec{e}_2\vec{e}_2
  + \frac{i\zeta_0 Wk_\perp\vt}{\omc}\vec{e}_1\times\mathbf{1}
  - \zeta_0^2 W\vec{e}_3\vec{e}_3.
\end{equation}
The argument of the $W$-function is $\zeta_0=\omega/|k_\parallel|\vt$.
Expanding the anisotropy tensor $\mathbf{Q}$ yields
\begin{equation}
  \mathbf{Q} \approx \mathbf{1} - \frac{q\Omega}{2\omc b_z}
  \left[1 - \frac{(8-q)\Omega}{4\omc b_z}\right]\ey\ey.
\end{equation}

\begin{figure}
  \centering\includegraphics{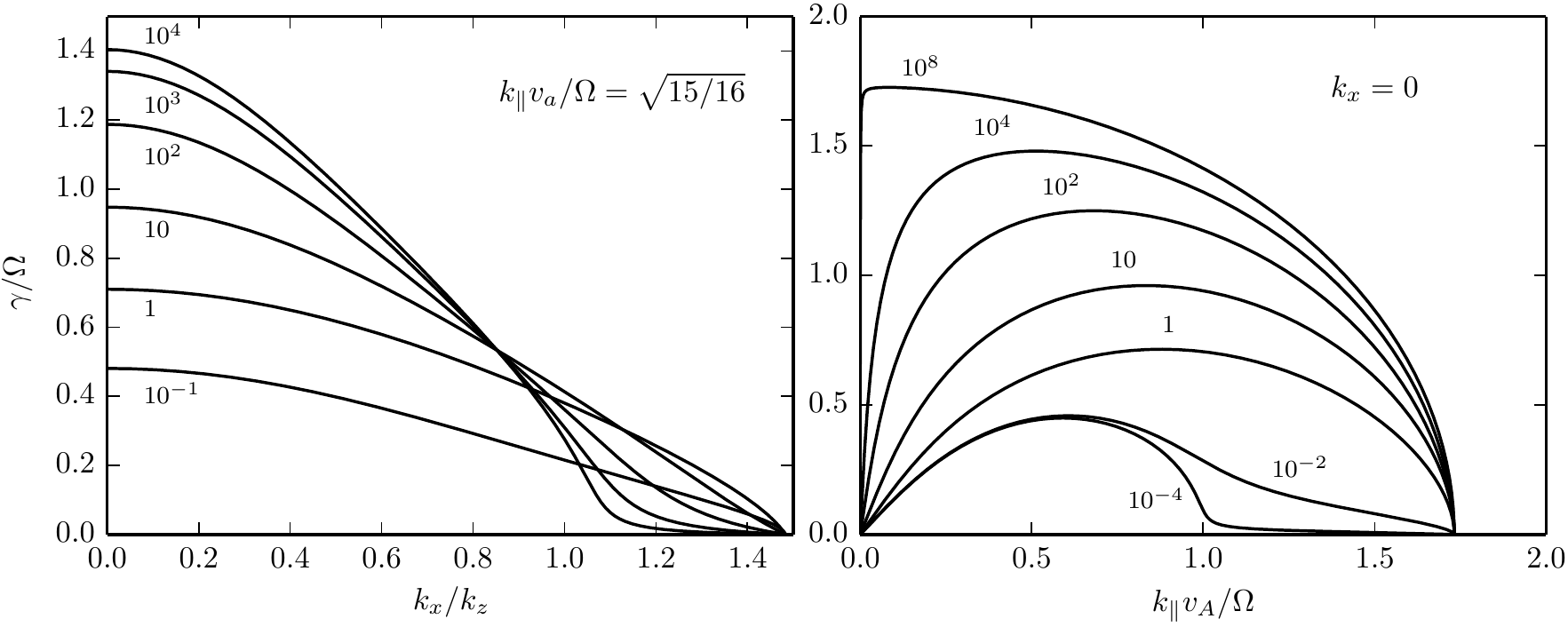}
  \caption{Dispersion relation of the MRI in the guiding center limit
    ($\omc/\Omega\to\infty$) for an inclined magnetic field with $b_z=b_y$.
    Labels give the value of $\beta_z=\beta/b_z^2$. These results agree well
    with the analogous plots shown in figs.~3 and 4 of
    \citet{Quataert2002}.}\label{fig:guiding-center}
\end{figure}

In order to derive the guiding center limit of the perpendicular dispersion
tensor given in \cref{eq:perpendicular-dispersion-single}, we need to compute
the tensor product $\mathbf{Q}\cdot\mathbf{\Lambda}\cdot\mathbf{Q}$ and
project the result onto the plane orthogonal to the magnetic field. Doing so
is complicated by the fact that for finite $e/m$ and $B_y\ne{}0$, the magnetic
field and the gyro-frequency vector $\vomg=\omg\vec{e}_3$ are misaligned, see
\cref{fig:stix}. This makes taking the limit $e/m\to\infty$ a tedious (albeit
straightforward) task. At the end of the day, the matrix representation of the
perpendicular dispersion tensor \labelcref{eq:perpendicular-dispersion-single}
in the basis $(\ex,\vec{b}\times\ex)$ is, however, given by the relatively
compact expression
\begingroup
\renewcommand*{\arraystretch}{1.4}
\begin{equation}
  \label{eq:DRguiding}
  \lim_{e/m\to\infty}\mathbf{D}_\perp \doteq
  \Bigg[\begin{matrix}
    k_\parallel^2\va^2 - \omega^2
    & 2i\omega\Omega b_z (1 + Wb_y^2/b_z^2) \\
    - 2i\omega\Omega b_z (1 + Wb_y^2/b_z^2)
    & k_\parallel^2\va^2 - \omega^2 - 2q\Omega^2 - 4\Omega^2b_y^2\zeta_0^2W
  \end{matrix}\Bigg]
  + \va^2\{1 + \beta(1-W)\}
  \Bigg[\begin{matrix}
    k_z^2 b_y^2 & k_x k_z b_y \\
    k_x k_z b_y & k_x^2
  \end{matrix}\Bigg],
\end{equation}
\endgroup
where $\beta=2\vt^2/\va^2$. \Cref{fig:guiding-center} shows plots of the
resulting dispersion relation $\det\mathbf{D}_\perp=0$ for different values of
$\beta_z=\beta/b_z^2$. The results are in good agreement with Figs.~3 and 4 of
\citet{Quataert2002}. The physics of the MRI in the guiding center limit are
discussed in detail in \citet{Quataert2002} \citep[see also][]{Balbus2004}.

\subsection{Parallel modes
  ($\vec{k}\bm{\parallel}\vec{B}\bm{\parallel}\vec{\Omega}$)}
\label{sec:par}

In this section we study modes whose wave vector is aligned with the magnetic
field, which in turn is aligned with the rotation axis. The ion response
tensor in this case is given by
\begin{equation}
  \label{eq:parallel-response}
  \mathbf{\Lambda} =
  \frac{1}{2}\sum_\pm\Lambda^{}_\pm
  (\vec{e}_1\pm i\vec{e}_2)(\vec{e}_1\mp i\vec{e}_2)
  - \frac{\omega^2}{k_z^2\vt^2} W\left(\frac{\omega}{|k_z|\vt}\right)
  \vec{e}_3\vec{e}_3,
\end{equation}
with perpendicular components
\begin{equation}
  \label{eq:circular-components}
  \Lambda_\pm = \frac{\omega}{\omega\pm\omg}
  \left[1 - W\left(\frac{\omega\pm\omg}{|k_z|\vt}\right)\right].
\end{equation}
At this point we note that even in the case of a purely vertical magnetic
field ($\vec{B}=B_z\ez$), the coordinate basis
$(\vec{e}_1,\vec{e}_2,\vec{e}_3)$ as defined in \cref{sec:plasma-response}
does not in general coincide with $(\ex,\ey,\ez)$. This is because
$\vec{e}_3=\vomg/\omg$, and whether the gyro-frequency vector $\vomg$ is
aligned or anti-aligned with the angular frequency vector
$\vec{\Omega}=\Omega\ez$ depends on the sign of $S_z=\omc{}b_z+2\Omega$. In
this context we also note that the gyration frequency $\omg>0$ throughout this
text.\footnote{Unless of course $\omg^2<0$, in which case much of our analysis
  is invalidated, see the discussion in \cref{sec:delta}.} The cyclotron
frequency $\omc=eB/m$ can have either sign, but this depends only on the sign
of the charge, \emph{not} on the direction of the magnetic field.

In the absence of Coriolis and tidal forces, $\Lambda_{+}$ ($\Lambda_{-}$)
would describe a right (left) handed circularly polarized wave. Insertion of
\cref{eq:parallel-response} with \cref{eq:circular-components} into
\cref{eq:perpendicular-dispersion-single} yields the perpendicular dispersion
tensor
\begin{equation}
  \label{eq:D-perp}
  \mathbf{D}_\perp \doteq
  k_z^2\va^2\begin{pmatrix} 1 & 0 \\ 0 & 1\end{pmatrix}
  + \omc^2\begin{pmatrix}
    1 & 0 \\ 0 & \omg/S_z
  \end{pmatrix}
  \cdot\begin{pmatrix}
    \Re & \Im \\ -\Im & \Re
  \end{pmatrix}
  \cdot\begin{pmatrix}
    1 & 0 \\ 0 & \omg/S_z
  \end{pmatrix}
  - i\omega\omc b_z\begin{pmatrix} 0 & -1 \\ 1 & 0\end{pmatrix}
  -2q\Omega^2\frac{\omc b_z}{S_z}\begin{pmatrix} 0 & 0 \\ 0 & 1\end{pmatrix},
\end{equation}
where the right hand side is the matrix representation of $\mathbf{D}_\perp$
in the basis $(\ex,\ey)$. We have also introduced the short hands
\begin{equation}
  \Re = \frac{1}{2}(\Lambda_{+} + \Lambda_{-})
  \quad\textrm{and}\quad
  \Im = \frac{1}{2i}(\Lambda_{+} - \Lambda_{-}).
\end{equation}
We note that $\Lambda_{-}=\Lambda_{+}^\ast$ for purely imaginary $\omega$. In
this case, $\Re$ and $\Im$ are simply the real and imaginary parts of
$\Lambda_{+}$.

\subsubsection{Cold ions}
\label{sec:cold-ions}

The dispersion relation simplifies significantly in the cold ion limit
$\vt\to0$. In this case we may drop the $W$-function in
\cref{eq:circular-components}. The dispersion relation
$\det\mathbf{D}_\perp=0$ may then be written as
\begin{equation}
  \label{eq:cold-mri-dispersion}
  \frac{k_z^4\va^4(\omega^2 - \omg^2)
  - k_z^2\va^2\omc b_z[q\Omega(\omega^2 - \kappa^2)
  - 2(\omega^2 + q\Omega^2)\omc b_z]
  - \omega^2(\omega^2 - \kappa^2)\omc^2}{\omega^2 - \omg^2} = 0.
\end{equation}
The numerator reproduces the Hall limit of the dispersion relation given by
\citet{Wardle1999}. We refer the reader to that paper as well as to
\citet{Balbus2001} for discussions of the Hall MRI\@. Here we only summarize
several key findings.

\Cref{eq:cold-mri-dispersion} is quadratic in both $\omega^2$ and $k^2$. Its
discriminant with respect to $\omega^2$ is positive definite. Thus there are
no over-stable solutions. The maximum growth rate is $\gamma=q\Omega/2$ at
\begin{equation}
  k_z = \frac{\Omega}{\va}
  \sqrt{1 - \alpha^4}
  {\left[1 + \frac{1}{2}(1 + \alpha^2)\frac{2\Omega}{\omc b_z}\right]}^{-1/2}
  \quad\textrm{with}\quad
  \alpha = \frac{\kappa}{2\Omega},
\end{equation}
cf.\ eq.~(29) of \citet{Wardle1999}. Remarkably, the maximum growth rate is
independent of $\omc$. The system is thus unstable even as $\omc\to0$. Below
we will see that this is also the case if the ions are hot. This behavior
might seem paradoxical in light of the result of \cref{sec:zero-mag}, where we
have shown that an initially unmagnetized charge-neutral cold collisionless
plasma is linearly stable in the shearing sheet. The resolution of this
apparent paradox is that here, it is only the ion cyclotron frequency
$\omc=\omega_{ci}$ that goes to zero whereas the electron cyclotron frequency
$\omega_{ce}$ is formally infinite since we have assumed the electrons to be
massless. We note, however, that for realistic plasmas with finite mass
ratios, it is at least questionable whether taking the limit
$\omega_{ce}\to\infty$ before letting $\omega_{ci}\to{}0$ is physically
meaningful.

\subsubsection{Gyro-viscous correction}
\label{sec:gyro-viscous}

In \cref{sec:guiding-center} we have shown that we recover the kinetic MHD
result obtained by \citet{Quataert2002} in the guiding center limit
$\omc\to\infty$. For parallel modes it is easy to calculate the first order
correction for finite cyclotron frequency.\footnote{Note that for parallel
  modes the MHD and kinetic MHD dispersion relations are identical} In order
to do so, let us assume the ordering
\begin{equation}
  \label{eq:ordering}
  \omega\sim k_z\va\sim\Omega\sim\epsilon\beta\Omega
  \sim\epsilon^2\omc,
\end{equation}
where the ordering parameter $\epsilon\ll1$. We thus assume that the plasma is
hot but not too hot in the sense that $1\ll\beta\ll\omc/\Omega$. The
perpendicular dispersion tensor to first order in $\epsilon$ is then given by
\begin{equation}
  \label{eq:gyro-viscous-drel}
  \mathbf{D}_\perp = (k_z^2\va^2 - \omega^2)(\mathbf{1} - \ez\ez)
  - i\omega\left(
    2\Omega - \frac{k_z^2\vt^2}{\omc b_z}
  \right)\ez\times\mathbf{1} - 2q\Omega^2\ey\ey.
\end{equation}
Setting its determinant equal to zero yields the dispersion relation
\begin{equation}
  \omega^4
  - \omega^2[\kappa^2 + 2k_z^2\va^2\left(1 - \beta\Omega/\omc b_z\right)]
  + k_z^2\va^2(k_z^2\va^2 - 2q\Omega^2) = 0.
\end{equation}
The lowest order correction to the standard MHD dispersion relation is thus of
order $\beta\Omega/\omc\ll{}1$. This is the magnitude of the gyro-viscous
force relative to the magnetic tension force in the Braginskii equations, a
fluid closure for weakly collisional plasmas that includes FLR
effects.\footnote{In this context we remark that the gyro-viscous force in the
  Braginskii equations does in fact \emph{not} depend on the collision
  frequency.} \citet{Ferraro2007} provides a numerical analysis of the MRI
with the gyro-viscous force taken into account. Our results confirm that
gyro-viscosity is indeed the dominant FLR effect in a hot plasma, not the Hall
effect.

Like the cold ion dispersion relation, \cref{eq:gyro-viscous-drel} is
quadratic in both $\omega^2$ and $k^2$ and may be analyzed in the same way.
The maximum growth rate
\begin{equation}
  \label{eq:maximum-growth-rate}
  \gamma = \frac{q\Omega}{2}{\left[
    1 + \frac{1}{2}(1-\alpha^4)\frac{\beta\Omega}{\omc b_z}
  \right]}^{1/2}
\end{equation}
occurs at
\begin{equation}
  \label{eq:optimal-wave-number}
  k_z = \frac{\Omega}{\va}\sqrt{1-\alpha^4}{\left[
    1 + \frac{1}{2}\frac{1-\alpha^8}{{(1+\alpha^2)}^2}
    \frac{\beta\Omega}{\omc b_z}
  \right]}^{1/2},
\end{equation}
where again $\alpha=\kappa/2\Omega$. As in the cold ion case there are no
over-stable solutions. 

\subsubsection{Numerical solution}

\begin{figure}
  \centering\includegraphics{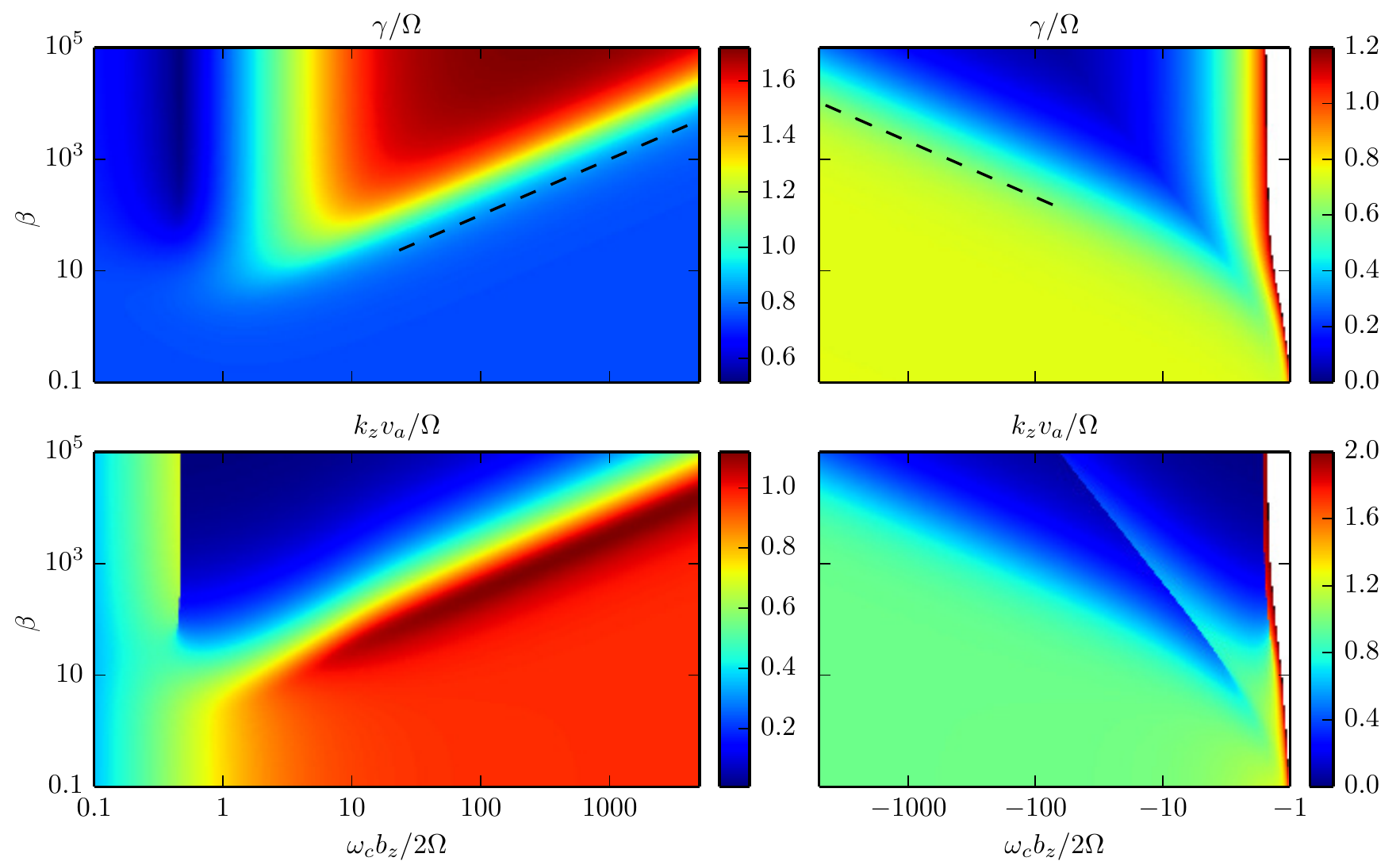}
  \caption{Growth rate and wavenumber of the fastest growing MRI mode for
    $\vec{k}\bm{\parallel}\vec{B}\bm{\parallel}\vec{\Omega}$ based on a
    numerical solution of the kinetic ion, cold, massless electron shearing
    sheet dispersion relation \labelcref{eq:D-perp}. The dashed black lines
    correspond to $\beta\Omega=\omc{}b_z$, the approximate ratio of the
    gyro-viscous force to the tension force.}\label{fig:vf-mri-2D}
\end{figure}

\begin{figure}
  \centering\includegraphics{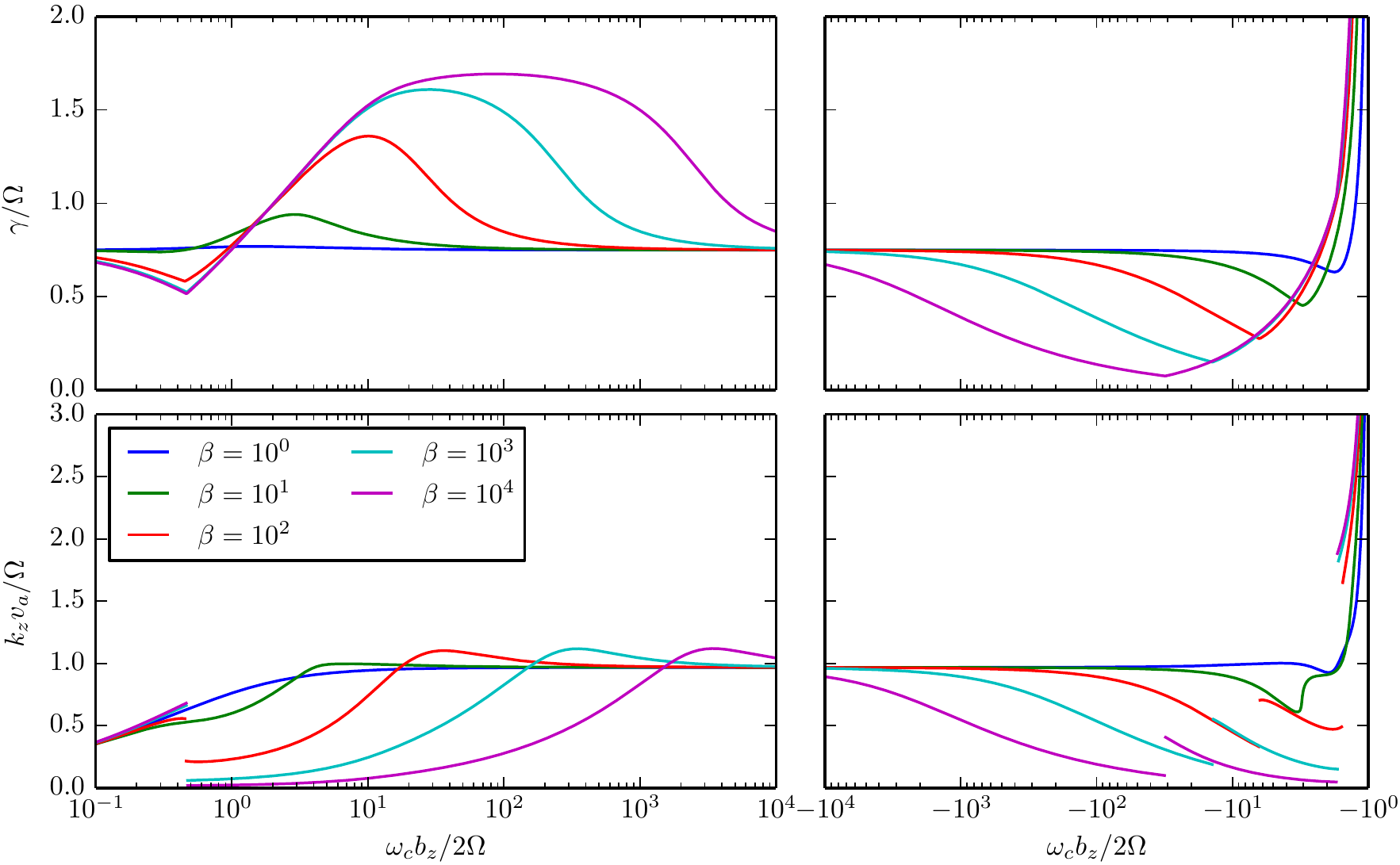}
  \caption{Growth rate and wavenumber of the fastest growing MRI mode as
    functions of $\omc{}b_z$ at fixed $\beta$. The cusps (upper panels) and
    discontinuities (lower panels) correspond to one local minimum of
    $\gamma(k_z)$ taking over another, see \cref{fig:drel}.}\label{fig:vf-mri}
\end{figure}

\begin{figure}
  \centering\includegraphics{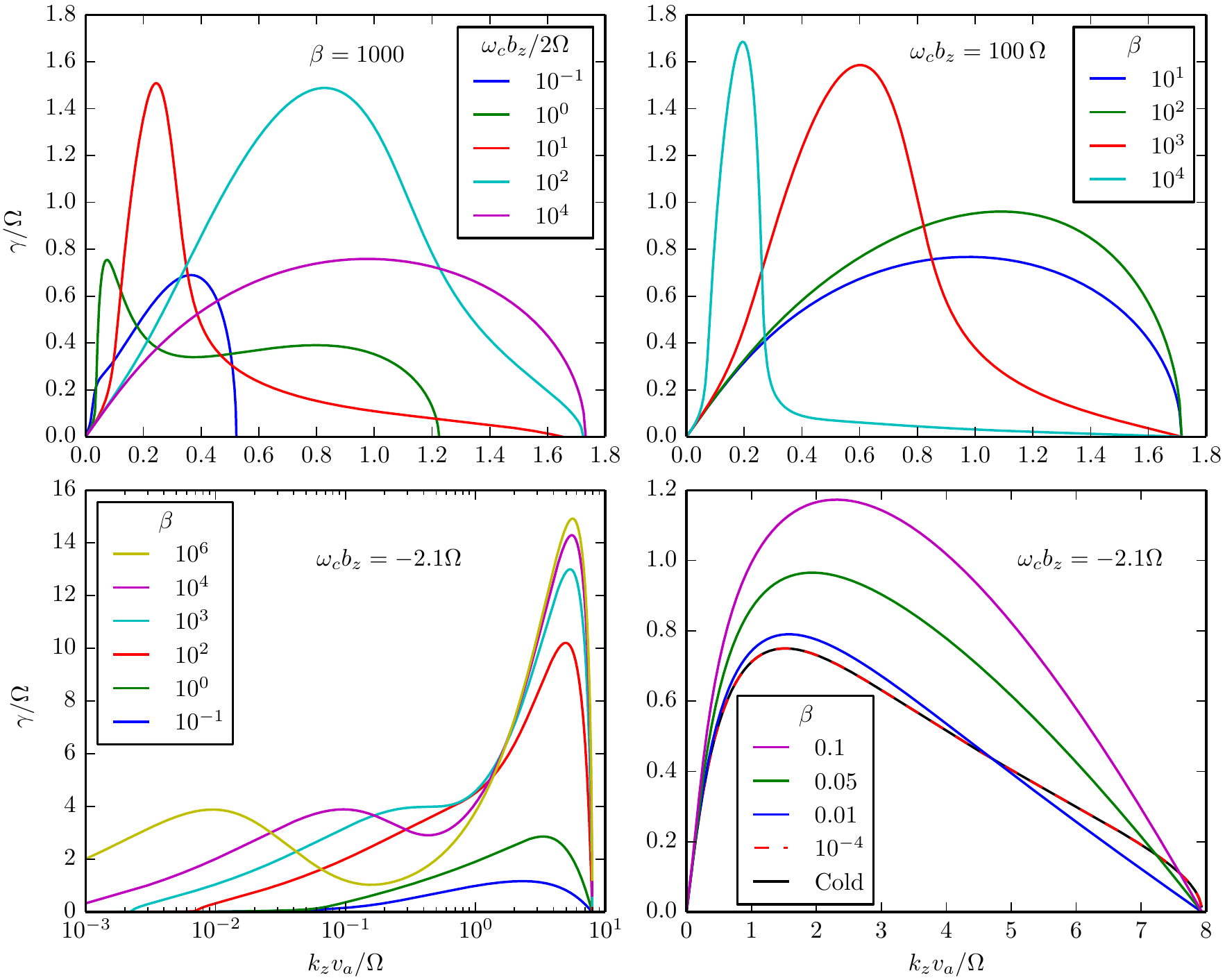}
  \caption{Plots of
    growth rate vs.\ wave number for select values of $\omc{}b_z$ and $\beta$.
    The lower panels illustrate the approach to the cold plasma dispersion
    relation as $\beta\to0$ when the tidal anisotropy is
    large.}\label{fig:drel}
\end{figure}

Outside of particular limits such as those discussed in
\cref{sec:cold-ions,sec:gyro-viscous} the dispersion relation even for the
simplest case of parallel modes with
$\vec{k}\bm{\parallel}\vec{B}\bm{\parallel}\vec{\Omega}$, given in
\cref{eq:D-perp}, is transcendental in $\omega$ and $k_z$ and its roots can in
general only be found numerically. This is the goal of this section.

Our numerical analysis has the caveat that we only look for purely growing
modes. In support of this we note that the MRI is known to be purely growing
in many fluid limits, including the cold ion/Hall MHD limit discussed in
\cref{sec:cold-ions}. In addition, the guiding center analysis carried out by
\citet{Quataert2002} also yielded purely growing modes only.\footnote{We
  stress that it would be unreasonable to assume that all unstable modes are
  purely growing. We make this assumption only for the MRI\@. Our general
  dispersion relation described by \cref{eq:dispersion-tensor} encompasses and
  extends the uniform plasma dispersion relation, and a uniform plasma is
  known to be subject to a host of instabilities that are not purely growing,
  particularly if said plasma is anisotropic with respect to the magnetic
  field \citep[see][and references therein]{Schekochihin2005,Sharma2006}.
  There is every reason to expect that these instabilities will carry over to
  the shearing sheet if the plasma is anisotropic with respect to the
  gyro-frequency vector, an anisotropy that our general dispersion relation
  allows for (see \cref{sec:steady-state}).}

For purely growing modes, the dispersion relation has a number of properties
that make the root finding particularly easy. First, because the conductivity
tensor vanishes as $\omega\to0$, the dispersion relation at marginal stability
is simply
\begin{equation}
  \label{eq:marginal-stability}
  k_z^2\va^2\left(k_z^2\va^2 - \frac{2q\Omega^2\omc}{S_z b_z}\right) = 0.
\end{equation}
Second, the dispersion tensor is real for purely growing modes, see
\cref{sec:par}. Thus, in order to find the roots of the dispersion relation we
may use Newton's method for real valued functions. In addition, it is easy to
get an overview of how many roots there are simply by plotting the dispersion
relation as a function of $\gamma=\sqrt{-\omega^2}$. Generally we find only a
single unstable root.

\Cref{eq:marginal-stability} determines the range of unstable wave numbers.
This range is finite provided that $\omc{}b_z>0$ or $\omc{}b_z<-2\Omega$, in
which case we only need to search for roots between $k_z=0$ and
\begin{equation}
  k_z = \frac{1}{\va}\sqrt{\frac{2q\Omega^2\omc}{S_z b_z}}.
\end{equation}
We do not consider the region of parameter space where
$-2\Omega\le\omc{}b_z<0$. In support of this we note that for
$-2\Omega\le\omc{}b_z\le(q-2)\Omega$ the tidal anisotropy $\Delta\ge1$, in
which case the equilibrium orbits are unbound and the shearing sheet
approximation rendered suspect, see the discussion in \cref{sec:delta}. For
$(q-2)\Omega<\omc{}b_z<0$ the cold ion dispersion relation has no unstable
roots \citep{Wardle1999} and we were not able to find any roots (visually) for
a hot plasma either.

We only consider Keplerian rotation ($q=3/2$) in our numerical solutions. The
free parameters are then $\omc{}b_z$ and $\beta$. We sweep through this
two-dimensional parameter space as follows. For each $\omc{}b_z$ we start with
$\beta\ll1$ and use the exact cold ion growth rate as our initial guess. We
then increase $\beta$ progressively and -- for each $k_z$ in the unstable
range -- use the previously obtained root as initial guess.

\Cref{fig:vf-mri-2D} shows pseudo color plots of the maximum growth rate and
associated wave number. Note that the color scales are not the same in the
left and right panels, because the MRI physics depends on the sign of $b_z$.
The black line in \cref{fig:vf-mri-2D} shows $\beta\Omega=\omc{}b_z$,
suggesting that gyro-viscosity is the dominant FLR effect throughout a large
portion of parameter space. \Cref{fig:vf-mri} shows cuts through
\cref{fig:vf-mri-2D} at fixed $\beta$. The cusps in the maximum growth rate
and the associated discontinuities in wave number are a consequence of the
dispersion relation having more than one local maximum, as we discuss below.

\Cref{fig:vf-mri} shows that for $\omc{}b_z\gtrsim\Omega$ the growth rate of
the fastest growing mode is always greater than the guiding center result
$\gamma=q\Omega/2$. The growth rate increases with $\beta$ and appears to
asymptote for a certain $\omc{}b_z\sim\beta\Omega$ to
$\gamma=\sqrt{2q}\Omega$. This is the same as the maximum growth rate obtained
by \citet{Quataert2002} and \citet{Balbus2004} without FLR effects but with a
45$^\circ$-inclined field; see also \cref{sec:guiding-center} and
\cref{fig:guiding-center}. For $0<\omc{}b_z\lesssim\Omega$ the system is less
unstable. As $\omc{}b_z\to0$ from above, the maximum growth and associated
wave number approach the cold ion result ($\gamma\to{}q\Omega/2$ and
$k_z\to0$). We note again, however, that this is a somewhat unphysical limit
in which the ion cyclotron frequency goes to zero while the electron cyclotron
frequency is formally infinite (recall that the latter is assumed throughout
this section).

The situation is different if the field is anti-aligned with the rotation
axis. For $\omc{}b_z\lesssim-4\Omega$ the system is always less unstable than
in the guiding center limit. As $\omc{}b_z\to-2\Omega$ from below, however,
the growth rate of the fastest growing mode as well as its wave number
diverge. While this might seem odd, we note that the tidal anisotropy also
diverges in this limit. This anisotropy evidently constitutes a free energy
reservoir that the system is able to tap into. We also note that the diverging
growth rate as $\omc{}b_z\to-2\Omega$ from below seems at odds with the cold
ion result ($\gamma=q\Omega/2$). This is only an apparent contradiction. The
growth rate does in fact approach the cold ion result for any
$\omc{}b_z<-2\Omega$ as $\beta\to0$.

\Cref{fig:drel} show plots of $\gamma(k_z)$ for various values of $\omc{}b_z$
and $\beta$. The dispersion relation can have more than one local maximum. The
discontinuities visible in the maximum growth rate and associated wavenumber
in \cref{fig:vf-mri-2D,fig:vf-mri} correspond to one local maximum in
\cref{fig:drel} overtaking another. The lower right panel demonstrates that
for a fiducial $\omc{}b_z=-2.1\Omega$, the growth rate indeed approaches the
cold ion result as $\beta\to0$.

\section{Conclusions}\label{sec:conclusions}

In this paper, we have studied the linear theory of a collisionless plasma in
the shearing sheet approximation. This approximation isolates the key
\emph{local} physics of differentially rotating disks and is free from the
complications of global boundary conditions. It is the ideal model for
studying the local dynamics of accretion disks. Our study is motivated by the
application to the magneto-rotational instability (MRI) in such disks if the
accreting plasma is collisionless.

We have for the first time derived the general linear conductivity tensor for
a collisionless plasma in the shearing sheet approximation. This accounts for
the complete linear dynamics of kinetic ions \emph{and} kinetic electrons with
finite Larmor radius/cyclotron frequency. Our calculation has leveraged the
extensive literature on analogous calculations for uniform plasmas
\citep{Ichimaru1973}. In particular, we have shown that by a suitable choice
of velocity space coordinates, the linear conductivity tensor for a
collisionless plasma in the shearing sheet approximation can be expressed
using standard results for a uniform plasma
(\cref{sec:conductivity-tensor,app:conductivity-tensor}). In doing so, our
primary assumptions are charge-neutrality, axisymmetric perturbations, and no
vertical stratification. The charge-neutral approximation is made both for its
applicability to realistic accretion disk dynamics and because the shearing
sheet approximation in its standard form relies on Galilean invariance, which
in turn necessitates charge-neutrality once the displacement current is
dropped \citep[see e.g.][]{Grad1966}.

The resulting dispersion relation for a collisionless plasma in the shearing
sheet approximation is given in \cref{eq:dispersion-tensor} with the
conductivity tensor given in \cref{eq:response-tensor-gyrotropic}. These
results apply for any equilibrium distribution function. In general, the tidal
force imposes a temperature anisotropy on the equilibrium distribution
function, see \cref{eq:tidal-anisotropy} and \cref{sec:steady-state}. This
anisotropy vanishes only in the guiding center limit in which the ratio of the
cyclotron frequency to the angular frequency $\omc/\Omega\rightarrow\infty$.
We denote a ``quasi-Maxwellian'' distribution function with this minimal level
of temperature anisotropy as the ``electromagnetic Schwarzschild
distribution'' because it is a generalization of the Schwarzschild
distribution of stellar dynamics to an ionized plasma. The conductivity tensor
for this distribution function is given in equations
\cref{eq:response-tensor-schwarzschild,eq:zeta-def,eq:T-tensor}. These results
for the linear theory of the shearing sheet are likely to be applicable to a
wide range of linear stability calculations, beyond those we have carried out
in this paper. 

We have focused on the application of our results to the linear theory of the
MRI in collisionless plasmas. In particular, we have shown that the extensive
existing literature on kinetic calculations of the MRI can all be understood
as approximations to the more general theory derived here. Specifically, the
guiding center kinetic MRI results of \citet{Quataert2002} follow from the
high cyclotron frequency and cold, massless electron limit of our dispersion
relation (\cref{sec:guiding-center}). The numerical gyro-viscous results of
\citet{Ferraro2007} -- which represent finite cyclotron frequency corrections
to the guiding center results of \citet{Quataert2002} -- follow analytically
from our dispersion relation by keeping the leading order cyclotron frequency
corrections to the guiding center result (\cref{sec:gyro-viscous}). And the
Hall MHD results of \citet{Wardle1999} -- which are particularly important in
the context of protostellar accretion disks -- follow from the cold ion and
cold, massless electron limit of our dispersion relation
(\cref{sec:cold-ions}).

In addition to providing a more general framework for understanding existing
calculations of the MRI in kinetic theory, we have also derived a number of
new results. In \cref{sec:cold-e,app:vlasov-fluid} we derive the dispersion
relation for the MRI in the limit of kinetic ions, but cold, massless
electrons. We solve this dispersion relation for the simplest case of
$\vec{k}\parallel\vec{\Omega}\parallel\vec{B}$ in \cref{sec:par}. These
numerical solutions of the full hot plasma dispersion relation confirm that
the primary finite cyclotron frequency correction to the guiding-center MRI
dispersion relation for $\beta\gtrsim1$ collisionless plasmas is not the Hall
effect, but rather gyro-viscosity \citep{Ferraro2007}.

The fluid electron, kinetic ion model of the shearing sheet studied in
\cref{sec:cold-e,app:vlasov-fluid} is known as the Vlasov-fluid model. It has
a number of attractive features for numerical studies of the MRI in
collisionless plasmas. In particular, the short length-scale, fast timescale
electron dynamics is ordered out allowing one to focus computational resources
on the ion dynamics that is likely the most important for the overall
evolution and saturation of the MRI\@.

As a second application of our results, we have derived the dispersion
relation of a cold plasma in the shearing sheet approximation
(\cref{sec:cold}). This problem was previously studied by \citet{Krolik2006},
who claimed that a cold, differentially rotating plasma that is initially
unmagnetized is linearly unstable to local axisymmetric perturbations. We have
shown, however, that such a plasma is in fact linearly stable
(\cref{sec:zero-mag}).

Starting with \cref{sec:par}, our analytical and numerical analysis of the MRI
in a hot plasma has focused on ``parallel modes'' with
$\vec{k}\parallel\vec{\Omega}\parallel\vec{B}$. In this case, the kinetic
corrections to the MRI (as well any other linear waves/modes that may be
present) are due to finite cyclotron frequency effects. However, the full
dispersion relation derived in \cref{sec:linear-theory} also includes finite
Larmor radius effects that become important when $k_\perp\vt\sim\omg$. While
differential rotation is unlikely to be important on such scales if
$\omc\gg\Omega$, the scale separation between $\omc$ and $\Omega$ that can be
achieved in numerical simulations is always limited (unless of course the
guiding center motion is averaged over, as in gyrokinetic simulations). Thus,
even if the interplay between orbital motion and Larmor motion is
insignificant in a given astrophysical system, our dispersion relation can be
useful for benchmarking a code used to simulate said system.

Possible future applications of our analysis include studies of the stability
of linear shear profiles in collisionless plasmas and the possibility of
velocity space instabilities driven by the background temperature anisotropy
present in differentially rotating plasmas.  In addition, all existing kinetic
calculations of the MRI assume cold electrons, motivated by the fact that hot
accretion flows onto compact objects are believed to have $T_i\gtrsim{}T_e$.
However, the general conductivity tensor given in
\cref{sec:conductivity-tensor} applies equally well to kinetic electrons and
can be used to assess the applicability of the cold electron approximation
under realistic accretion disk conditions.  Finally, our calculations can be
used to determine the properties of the MRI under the conditions achievable in
laboratory experiments aimed at studying low collisionality high-$\beta$
plasmas \citep[e.g.][]{Collins2012}.

\acknowledgments{}

We thank Omer Blaes, Martin Pessah, Mario Riquelme, Anatoly Spitkovsky, and
Scott Tremaine for useful conversations. In particular, conversations with
Anatoly Spitkovsky inspired us to think in more detail about the unmagnetized
limit (\cref{sec:zero-mag}). We also thank Ellen Zweibel and Julian Krolik for
a free exchange of ideas that prompted us to sharpen our understanding of the
cold plasma dispersion relation (\cref{sec:cold}). This work was supported in
part by NASA HTP grant NNX11AJ37G, NSF grants PHY11--25915 and AST-1333682, a
Simons Investigator award from the Simons Foundation, and the Thomas Alison
Schneider Chair in Physics at UC Berkeley.

\appendix
\crefalias{section}{appsec}

\section{The unperturbed Hamiltonian}
\label{app:unperturbed-hamiltonian}

In this section we derive the unperturbed Hamiltonian given in
\cref{eq:hamiltonian}. Let us start with the Hamiltonian governing the
dynamics of charged particles in a rotating frame. This is given by
\begin{equation}
  \mathcal{H} = \frac{1}{2}{\left(
    \vec{p} - \vec{\Omega}\times\vec{r} - \cm\vec{A}
  \right)}^2 + \psi + \cm\Phi,
\end{equation}
where $\vec{A}$ is the magnetic vector potential and $\Phi$ is the
electrostatic potential. The coordinate system is oriented such that
$\vec{\Omega}=\Omega\ez$. In the shearing sheet approximation, the tidal
potential $\psi$ is given as in \cref{eq:tidal-potential} by
\begin{equation}
  \psi = -q\Omega^2 x^2 + \frac{1}{2}\vfreq^2 z^2.
\end{equation}
It is straightforward to verify that
$\partial{}f/\partial{}t+\{f,\mathcal{H}\}=0$ is the same as the Vlasov
equation \labelcref{eq:vlasov-sheet} when the former is expressed in terms of
the particle velocity $\vec{v}=\vec{p}+\vec{\Omega}\times\vec{r}+e\vec{A}/m$.

We now consider the equilibrium discussed in \cref{sec:steady-state},
consisting of a linear shear flow and a uniform magnetic field. Without loss
of generality we may (at first) work in the so-called symmetric gauge, in
which the vector potential is given by $\vec{A}=\vec{B}\times\vec{r}/2$. The
magnetic field is assumed frozen into the background shear flow. The
corresponding electric field derives from the electrostatic potential
$\Phi=-q\Omega{}xA_y$. With this, the Hamiltonian may be written as
\begin{equation}
  \label{eq:shearing-sheet-hamiltonian}
  2\mathcal{H} = {\left(
    \vec{p} - \frac{\vec{S}\times\vec{r}}{2}
  \right)}^2 - q\Omega S_z x^2 + \vfreq^2 x^2.
\end{equation}
The Hamiltonian \labelcref{eq:hamiltonian} is obtained after a canonical
transformation to new coordinates $(\vec{r}',\vec{p}')$ derived from the
type-2 generating function
\begin{equation}
  G(\vec{r},\vec{p}') = xp_x' + yp_y' - S_z xy/2.
\end{equation}

\section{Global equilibrium}
\label{app:global}

In this section we investigate the stability of charged particles on circular
orbits in a global disk equilibrium. We also derive a generalization of the
modified Schwarzschild distribution \citep{Shu1969} and show that for nearly
circular orbits it reduces to the ``electromagnetic Schwarzschild
distribution'' used in the main text.

\subsection{Bulk flow and electromagnetic field configuration}

We seek an axisymmetric equilibrium with a purely rotational bulk flow common
to all plasma species. In a cylindrical coordinate system $(r,\varphi,z)$,
this bulk flow may be written as
\begin{equation}
  \vec{u} = r\Omega\nabla\varphi,
\end{equation}
where $\Omega(r,z)$ is the orbital frequency. Following
\citet{Kaufman1972,Cary1983} we write the magnetic field as
\begin{equation}
  \label{eq:global-bfield}
  \vec{B} =
  \nabla w(\psi)\times\nabla\vartheta - \nabla\psi\times\nabla\varphi,
\end{equation}
where $\psi(r,z)$ is the poloidal flux function. The poloidal angle
$\vartheta(r,z)$ may be chosen to increase clockwise from $-\pi$ to $\pi$
along poloidal field lines, which are isocontours of $\psi$. Note that the
first term on the right hand side of \cref{eq:global-bfield} represents the
toroidal field. The second term is the usual parametrization of an
axisymmetric poloidal field. The magnetic field given in
\cref{eq:global-bfield} derives from the axisymmetric vector potential
\begin{equation}
  \vec{A} = w(\psi)\nabla\vartheta - \psi\nabla\varphi.
\end{equation}

The electric field is determined from the condition that the magnetic field be
frozen into the equilibrium flow, to wit
\begin{equation}
  \vec{E} = -\vec{u}\times\vec{B}.
\end{equation}
This must be irrotational, thus
\begin{equation}
  \nabla\times\vec{E} =
  \left(
    \pder{\vartheta}{r}\pder{\psi}{z} -
    \pder{\psi}{r}\pder{\vartheta}{z}
  \right)
  \pder{\Omega}{\vartheta}r\nabla\varphi = 0.
\end{equation}
The factor in parentheses is the Jacobian of the transformation from flux
coordinates $(\psi,\varphi,\vartheta)$ to cylindrical coordinates
$(r,\varphi,z)$. Assuming this is non-singular, it follows that the orbital
frequency $\Omega(\psi)$ is constant on poloidal field lines
\citep{Ferraro1937}, see also \citet{Papaloizou1992}. With this result, it
easy to show that the electric field derives from the electrostatic potential
\begin{equation}
  \Phi(\psi) = -\int\!\Omega(\psi)\,d\psi.
\end{equation}

\subsection{Stability of circular orbits}

The equations of motion for charged particles orbiting in the equilibrium
fields are generated by the Hamiltonian
\begin{equation}
  \label{eq:global-hamiltonian}
  \mathcal{H} = \frac{1}{2}\left[
    {\left(p_r - \cm A_r\right)}^2
    + \frac{1}{r^2}{\left(p_\varphi + \cm\psi\right)}^2
    + {\left(p_z - \cm A_z\right)}^2
  \right] + \Psi + \cm\Phi,
\end{equation}
where $\Psi(r,z)$ is the gravitational potential. The equilibrium Hamiltonian
is independent of $\varphi$ due to the assumed axial symmetry. The canonical
angular momentum
\begin{equation}
  \label{eq:canonical-angular-momentum}
  p_\varphi = r^2\dot{\varphi} - \cm\psi
\end{equation}
is thus an integral of motion and for a given $p_\varphi$, the dynamics may be
described in the reduced phase space $(r,z,p_r,p_z)$.

We assume that the equilibrium also has equatorial symmetry. By this we mean
that there is a plane of symmetry at $z=0$ (the \emph{mid-plane}) and that the
equilibrium fields are either even or odd functions of $z$. Specifically, we
take the gravitational and electrostatic potentials as well as the poloidal
flux $\psi$ to be even, and the poloidal angle $\vartheta$ to be odd. This
completely specifies the equatorial symmetry of the equilibrium.

Using Hamilton's equation, the first variation of \cref{eq:global-hamiltonian}
is $\delta\mathcal{H}=\dot{r}\delta{}p_r+\dot{z}\delta{}p_z
-\dot{p}_r\delta{}r-\dot{p}_z\delta{}z$. All equilibrium points in phase space
(for which $\delta\mathcal{H}=0$) are thus circular orbits
($\dot{r}=\dot{z}=0$). For such orbits, the rates of change of the canonical
momenta are
\begin{align}
  \dot{p}_r &= \cm r(\dot{\varphi} - \Omega) B_z
  - \pder{\Psi}{r} + r\dot{\varphi}^2 \\
  \dot{p}_z &= \cm r(\Omega - \dot{\varphi}) B_r
  - \pder{\Psi}{z},
\end{align}
where we have made use of our assumption that the electric field is frozen
into the bulk flow $\vec{u}=r^2\Omega\nabla\varphi$. Given the equatorial
symmetry of $\Psi$, particles on circular orbits are (relative) equilibria if
they travel in the mid-plane ($z=0$) and their angular velocity matches that
of the bulk flow, i.e.\ if
\begin{equation}
  \dot{\varphi} = \Omega(r,0),
\end{equation}
provided the latter arises due to gravito-centrifugal balance, i.e.\ provided
that
\begin{equation}
  r\Omega^2(r,0) = \pder{\Psi(r,0)}{r}.
\end{equation}
From now on it will be understood that all quantities are evaluated at $z=0$.

In order to determine the stability of circular orbits we can invoke the
so-called Lagrange-Dirichlet theorem \citep[see e.g.][and references
therein]{Krechetnikov2007}. This states that an equilibrium point (in this
case the circular orbit) is stable to finite perturbations (nonlinear
stability) if the second variation of $\mathcal{H}$ (a quadratic form) or,
equivalently, its Hessian is positive definite. In the phase space coordinate
system $(r,z,p_r,p_z)$, the Hessian has the matrix representation
\begingroup
\renewcommand*{\arraystretch}{2.5}
\begin{equation}
  \label{eq:hessian}
  \mathbf{D} \doteq
  \begin{bmatrix}
    \displaystyle S_z\left(S_z + r\pder{\Omega}{r}\right)
    + {\left(\cm\pder{A_z}{r}\right)}^2 & 0 & 0
    & \displaystyle -\cm\pder{A_z}{r}
    \\
    0 & \displaystyle\pdder{\Psi}{z} + {\left(\cm\pder{A_r}{z}\right)}^2
    & \displaystyle -\cm\pder{A_r}{z} & 0 \\
    0 & \displaystyle -\cm\pder{A_r}{z} & 1 & 0 \\
    \displaystyle -\cm\pder{A_z}{r} & 0 & 0 & 1
  \end{bmatrix},
\end{equation}
\endgroup
where
\begin{equation}
  \label{eq:global-spin}
  \vec{S} = \cm\vec{B} + 2\vec{\Omega}
\end{equation}
is defined in the same way as in the main text, see \cref{eq:spin}. The
Hessian matrix given in \cref{eq:hessian} is positive definite if and only if
\begin{equation}
  \label{eq:positive-definite}
  S_z\left(S_z + r\pder{\Omega}{r}\right) > 0
  \quad\textrm{and}\quad
  \pdder{\Psi}{z} > 0.
\end{equation}
Otherwise the Hessian is indefinite. If the above inequalities are satisfied,
then circular orbits minimize the energy. Note that if we define
\begin{equation}
  \label{eq:global-tidal-anisotropy}
  \Delta = -\frac{r}{S_z}\pder{\Omega}{r},
\end{equation}
in complete analogy to \cref{eq:tidal-anisotropy}, then the first inequality
in \cref{eq:positive-definite} is equivalent to $\Delta<1$. Note also that for
circular orbits, the radial derivative of the canonical angular momentum is
given by
\begin{equation}
  \label{eq:angular-momentum-gradient}
  \pder{p_\varphi}{r} = r\left(S_z + r\pder{\Omega}{r}\right) =
  rS_z(1 - \Delta).
\end{equation}

For neutral particles, the condition for stability $\Delta<1$ reduces to the
familiar Rayleigh criterion
\begin{equation}
  \label{eq:global-rayleigh-criterion}
  \kappa^2 = 2\Omega\left(2\Omega + r\pder{\Omega}{r}\right) > 0.
\end{equation}
From \cref{eq:angular-momentum-gradient} we recover the familiar result
\citep[e.g.][]{Chandrasekhar1961} that the angular momentum is a decreasing
function of radius in this case. For charged particles, the (canonical)
angular momentum in a stable configuration does not necessarily decrease with
radius because $S_z$ is not sign definite.

\subsubsection{Linear stability}

Above we have shown that circular orbits are nonlinearly stable provided that
$\Delta<1$ and $\partial^2\Psi/\partial{}z^2>0$. This does not necessarily
imply that circular orbits are \emph{unstable} if $\Delta>1$.\footnote{We use
  the terms linear and nonlinear stability rather loosely. For more precise
  definitions see e.g.\ \citet{Holm1985}.} To see this, we consider the
linearized Hamiltonian system
\begin{equation}
  \delta\dot{\vec{w}} = \mathbf{J}\cdot\mathbf{D}\cdot\delta\vec{w},
\end{equation}
where $\vec{w}=(r,z,p_r,p_z)$,
\begin{equation}
  \mathbf{J} = \begin{bmatrix}
     0 &  0 & 1 & 0 \\
     0 &  0 & 0 & 1 \\
    -1 &  0 & 0 & 0 \\
     0 & -1 & 0 & 0
  \end{bmatrix},
\end{equation}
and $\mathbf{D}$ is the Hessian given in \cref{eq:hessian}. Assuming that
$\delta\vec{w}$ depends on time as $\exp(-i\omega{}t)$, the linearized system
has non-trivial solutions provided that
\begin{equation}
  \label{eq:solubility-condition}
  \det(i\omega\mathbf{1} + \mathbf{J}\cdot\mathbf{D}) =
  \omega^4 - \omega^2\left[
    S_z\left(S_z + r\pder{\Omega}{r}\right) + S_\varphi^2 + \pdder{\Psi}{z}
  \right]
  + S_z\left(S_z + r\pder{\Omega}{r}\right)\pdder{\Psi}{z} = 0,
\end{equation}
where $\mathbf{1}$ denotes the identity matrix.

Let us first consider the case of a vanishing toroidal magnetic field. For
$S_\varphi=eB_\varphi/m=0$, the solubility condition
\labelcref{eq:solubility-condition} dramatically simplifies because the radial
and vertical degrees of freedom decouple. The stability of the respective
motions is determined by
\begin{equation}
  \omega^2 = S_z\left(S_z + r\pder{\Omega}{r}\right)
  \quad\textrm{and}\quad
  \omega^2 = \pdder{\Psi}{z}.
\end{equation}
Comparison with \cref{eq:positive-definite} shows that circular orbits are
linearly stable if they are nonlinearly stable (i.e.\ if the minimize the
energy), and linearly unstable otherwise.

The situation is more complicated if there is a toroidal magnetic field. In
this case there is a region of parameter space where circular orbits are
linearly stable even though they do not minimize the energy. To see this, let
us assume (as in the bulk of the main text) that there is no stratification
($\Psi=\mathrm{const}$). In this case, the solubility condition
\cref{eq:solubility-condition} reduces to
\begin{equation}
  \label{eq:solubility-no-stratification}
  \omega^2(\omega^2 - \omg^2) = 0,
\end{equation}
where the square of the gyration frequency $\omg$ is given by
\begin{equation}
  \label{eq:global-gyration-freq}
  \omg^2 = S_z\left(S_z + r\pder{\Omega}{r}\right) + S_\varphi^2.
\end{equation}
This is the global analogue of the local gyration frequency defined in
\cref{eq:gyro-frequency} in the main text.
\Cref{eq:solubility-no-stratification} with \cref{eq:global-gyration-freq}
implies that circular orbits are linearly stable provided that
\begin{equation}
  \Delta < 1 + S_\varphi^2/S_z^2,
\end{equation}
which does \emph{not} coincide with the nonlinear stability boundary
$\Delta<1$. In this context we remark that according to a theorem due to
\citet{Bloch1994}, a relative equilibrium\footnote{Circular orbits are
  relative equilibria.} that is linearly but \emph{not} nonlinearly stable
becomes linearly unstable when a small amount of dissipation is added to the
system. We are thus led to conclude that circular orbits in realistic systems
are always unstable if $\Delta>1$, regardless of whether or not there is a
toroidal magnetic field.

\subsection{The modified electromagnetic Schwarzschild distribution}

The electromagnetic Schwarzschild distribution \labelcref{em:schwarzschild}
may be viewed as a generalization of the Schwarzschild distribution
\citep[e.g.][]{Julian1966}. The latter is the local limit (nearly circular
orbits) of the \emph{modified} Schwarzschild distribution as derived by
\citet{Shu1969}. Here we generalize the modified Schwarzschild distribution
and show that it reduces to \cref{em:schwarzschild} in the local limit.

The integrals of motion are the particle energy
\begin{equation}
  \mathcal{E} = \frac{\dot{r}^2 + r^2\dot{\varphi}^2 + \dot{z}^2}{2}
  + \Psi + \cm\Phi
\end{equation}
and the canonical angular momentum $p_\varphi$ defined in
\cref{eq:canonical-angular-momentum}.  By Jeans' theorem, any distribution
function of the form $f(\mathcal{E},p_\varphi)$ is a solution of the steady
state Vlasov equation. Let us define the guiding center radius $\rc$ through
\begin{equation}
  p_\varphi = \rc^2\Omega(\rc,0) - \cm\psi(\rc,0).
\end{equation}
Thus, $\rc$ is the radius of circular orbit with angular momentum $p_\varphi$.
Its energy is
\begin{equation}
  \mathcal{E}_c = \frac{\rc^2\Omega^2(\rc,0)}{2}
  + \Psi(\rc,0) + \cm\Phi(\rc,0).
\end{equation}

Note that $\mathcal{E}_c$ depends only on the angular momentum and is thus an
integral of motion. We define the modified electromagnetic Schwarzschild
distribution as
\begin{equation}
  \label{eq:modified-em-schwarzschild}
  f \propto \exp(-\mathcal{K}/T),
\end{equation}
where
\begin{equation}
  \label{eq:global-gyration-energy}
  \mathcal{K} = \mathcal{E} - \mathcal{E}_c
\end{equation}
is the gyration energy. The proportionality constant in
\cref{eq:modified-em-schwarzschild} and the temperature $T$ may in general
depend on the canonical angular momentum.

We now show that for nearly circular orbits,
\cref{eq:modified-em-schwarzschild} reduces to the electromagnetic
Schwarzschild distribution given in \cref{em:schwarzschild}. Let us first
introduce the peculiar velocity
\begin{equation}
  \tvec{v} = \dot{\vec{r}} - r\Omega(r,z)\vec{e}_\varphi.
\end{equation}
Substituting
\begin{equation}
  r\dot{\varphi} = \frac{1}{r}\left[p_\varphi + \cm\psi(r,z)\right] =
  \frac{1}{r}\left[\rc^2\Omega(\rc,0) + \cm\psi(r,z) - \cm\psi(\rc,0)\right]
\end{equation}
and expanding for small $\tilde{v}_\varphi/r\Omega$, the azimuthal component
of the peculiar velocity is approximately given by
\begin{equation}
  \label{eq:peculiar-velocity}
  \tilde{v}_\varphi =
  \left[S_z(r,0) + r\pder{\Omega(r,0)}{r}\right](r - \rc),
\end{equation}
where $\vec{S}$ is defined in \cref{eq:global-spin}. Corrections to
\cref{eq:peculiar-velocity} are quadratic in $r-\rc$ and $z$. Expanding the
gyration energy defined in \cref{eq:global-gyration-energy} to quadratic order
in $\tilde{v}_\varphi/r\Omega$ yields
\begin{equation}
  \label{eq:global-gyration-energy-approx}
  2\mathcal{K} = \tilde{v}_r^2
  + S_z(r,0)\left[S_z(r,0) + r\pder{\Omega(r,0)}{r}\right]
  {(r - \rc)}^2 + \tilde{v}_z + \nu^2 z^2,
\end{equation}
Here, the square of the vertical frequency $\nu$ is defined as
\begin{equation}
  \nu^2 = \pdder{\Psi}{z}
\end{equation}
evaluated at $z=0$. Combining \cref{eq:peculiar-velocity} into
\cref{eq:global-gyration-energy-approx} finally yields
\begin{equation}
  \mathcal{K} = \frac{1}{2}\left[
    \tilde{v}_r^2 + \frac{\tilde{v}_\varphi^2}{1 - \Delta} + \tilde{v}_z^2
  \right] + \frac{\nu^2 z^2}{2},
\end{equation}
which is just the electromagnetic Schwarzschild distribution defined in
\cref{eq:gyration-energy}.

\section{The conductivity tensor}
\label{app:conductivity-tensor}

Here we fill in some of the details of going from the formal solution to the
linearized Vlasov equation, \cref{eq:formal-solution}, to the ``conductive''
relationship between current and electric field,
\cref{eq:conductive-relation}. Note that we are assuming axisymmetric
perturbations, so $k_y=0$.

Using Faraday's law in the form of \cref{eq:linearized-faraday}, the formal
solution is given by
\begin{equation}
  \label{eq:formal-solution-app}
  \delta f_s = -\frac{1}{i\omega}\frac{\qs}{m_s}\int_{-\infty}^t\!dt'\,
  \pder{f_s}{\tvec{v}'}
  \cdot[i(\omega - \vec{k}\cdot\tvec{v}')\mathbf{1} + i\vec{k}\tvec{v}'
  + q\Omega\ex\ey]
  \cdot\delta\tvec{E},
\end{equation}
where $\vec{r}'(t')$ and $\tvec{v}'(t')$ satisfy
$d\vec{r}'/dt'=\tvec{v}'-q\Omega{}x'\ey$ and
\begin{equation}
  \label{eq:unperturbed-eom}
  \der{\tvec{v}'}{t'} = \frac{\qs}{m_s}\tvec{v}'\times\vec{B}
  - 2\vec{\Omega}\times\tvec{v}' + q\Omega\tilde{v}_x'\ey
\end{equation}
subject to the ``final conditions'' $\vec{r}'(t)=\vec{r}$ and
$\tvec{v}'(t)=\tvec{v}$. Taking the first moment of
\cref{eq:formal-solution-app}, multiplying by $\qs n_s$, and summing over
species yields the perturbed current density
\begin{equation}
  \label{eq:conductive-relation-tilde}
  \delta\vec{J} = \tvec{\sigma}
  \cdot\left(\mathbf{1} - \frac{q\Omega}{i\omega}\ex\ey\right)
  \cdot\delta\tvec{E}.
\end{equation}
Here, we have introduced the tensor
\begin{equation}
  \label{eq:sigma-tilde}
  \tvec{\sigma}
  = -\frac{1}{i\omega}\sums\frac{\qs^2n_s}{m_s}\int d^3v\int_0^\infty d\tau\,
  \tvec{v}\pder{f_s}{\tvec{v}'}
  \cdot[i(\omega - \vec{k}\cdot\tvec{v}')\mathbf{1} + i\vec{k}\tvec{v}'
  + q\Omega\ex\ey]\exp\{-i\phi(\tau)\},
\end{equation}
where the new time variable $\tau=t-t'$ and the short hand
\begin{equation}
  \phi(\tau) = \vec{k}\cdot(\vec{r} - \vec{r}') - \omega\tau.
\end{equation}
From now on we will suppress the species index when there is no risk of
confusion. In order to express \cref{eq:sigma-tilde} in terms of the
``gyrotropic velocity'' $\vvc=\mathbf{Q}^{-1}\cdot\vec{v}$, let us first note
that
\begin{equation}
  \label{eq:df-dvtwiddle}
  \pder{f}{\tvec{v}}
  = \pder{f}{\vvc}\cdot\mathbf{Q} + \Delta\pder{f}{\tilde{v}_y}\ey,
\end{equation}
where the anisotropy tensor $\mathbf{Q}$ is defined in
\cref{eq:anisotropy-tensor}. With this, the orbit integral in
\cref{eq:sigma-tilde} may be rewritten as
\begin{multline}
  \label{eq:by-parts}
  \int_0^\infty d\tau\,\pder{f}{\tvec{v}'}
  \cdot[i(\omega - \vec{k}\cdot\tvec{v}')\mathbf{1} + i\vec{k}\tvec{v}'
  + q\Omega\ex\ey]
  \,\exp\{-i\phi(\tau)\} \\
  = -\Delta\pder{f}{\tilde{v}_y}\ey
  + \int_0^\infty d\tau\,\pder{f}{\vvc'}
  \cdot[i(\omega - \vec{k}\cdot\vvc')\mathbf{1} + i\vec{k}\vvc']
  \cdot\mathbf{Q}\exp\{-i\phi(\tau)\}.
\end{multline}
In obtaining this result we have used that
\begin{equation}
  \label{eq:some-identity}
  q\Omega\pder{f}{\tilde{v}_x'}
  - \Delta\pder{}{\tau}\pder{f}{\tilde{v}_y'}
  = \left(
    q\Omega\tilde{v}_x' - \frac{\Delta}{1-\Delta}\pder{\tilde{v}_y'}{\tau}
  \right)
  \frac{1}{\vc_\perp}\pder{f}{\vc_\perp} = 0.
\end{equation}
The second equality in follows from \cref{eq:unperturbed-eom}. To get the
first equality, notice that
\begin{equation}
  \pder{f}{\tilde{v}_x} = \frac{\tilde{v}_x}{\vc_\perp}\pder{f}{\vc_\perp}
  \quad\textrm{and}\quad
  \pder{f}{\tilde{v}_y} =
  \frac{1}{1-\Delta}\frac{\tilde{v}_y}{\vc_\perp}\pder{f}{\vc_\perp}
  + \frac{\vc_\parallel\sin\varphi}{\sqrt{1-\Delta}}\left(
    \frac{1}{\vc_\perp}\pder{f}{\vc_\perp}
    - \frac{1}{\vc_\parallel}\pder{f}{\vc_\parallel}
  \right),
\end{equation}
where $\varphi$ is defined through $\vomg=\ez\cos\varphi-\ey\sin\varphi$. The
second term in the expression for $\partial{}f/\partial\tilde{v}_y$ does not
depend on time and therefore makes no contribution in \cref{eq:some-identity}.
Using \cref{eq:by-parts} in \cref{eq:sigma-tilde} and integrating the
contribution from the first term on the right hand side of \cref{eq:by-parts}
by parts yields
\begin{equation}
  \label{eq:sigma-twiddle}
  \tvec{\sigma} = \vec{\sigma}
  - \frac{1}{i\omega}\sums\frac{\qs^2n_s}{m_s}\Delta_s\ey\ey
\end{equation}
with the conductivity tensor $\vec{\sigma}$ as defined in
\cref{eq:conductivity}. Inserting \cref{eq:sigma-twiddle} into
\cref{eq:conductive-relation-tilde} then finally yields
\cref{eq:conductive-relation}.

\section{Vlasov-fluid dispersion relation}
\label{app:vlasov-fluid}

The Vlasov-fluid equations \citep{Freidberg1972} describe the low-frequency
electromagnetic interaction of collisionless ions with a massless and
isotropic electron fluid. A detailed description is given in the recent work
by \citet{Cerfon2011}. Here we show explicitly that the Vlasov fluid
dispersion relation in the shearing sheet is equivalent to the cold, massless
electron limit of the general hot plasma shearing sheet dispersion relation
derived in \cref{sec:cold-e} (up to a term related to electron pressure that
is included in the Vlasov fluid model as a crude representation of finite
electron temperature effects). The Vlasov fluid calculation presented here is
useful primarily in that it provides an independent formalism for calculating
the shearing sheet dispersion relation in the limit in which ion dynamics are
the most important. In addition, the Vlasov fluid derivation is somewhat more
similar to the MHD calculation in that it utilizes the Lagrangian displacement
and the fact that field lines are frozen into the electron fluid.

\subsection{The Vlasov-fluid equations}

The electrons dynamics are described by the momentum equation
\begin{equation}
  \label{eq:ohms-law}
  e n_e(\vec{E} + \ue\times\vec{B}) + \nabla\pe = 0,
\end{equation}
which we will henceforth refer to as the generalized Ohm's law. Here, we
assume that the electron pressure $\pe$ is a function of the electron number
density $n_e$ alone, i.e.\ we assume a barotropic equation of state.
\Cref{eq:ohms-law} arises out of the electron momentum equation in the limit
of vanishing electron mass and assuming an isotropic pressure. The electrons
must be collisional for this assumption to be reasonable. Indeed,
\citet{Rosin2011} derive \cref{eq:ohms-law} rigorously through an expansion in
the electron-to-ion mass ratio under the assumption of weakly collisional ions
but strongly collisional electrons.

The crude electron model represents the most severe limitation of Vlasov-fluid
theory. At the same time it may, however, also be viewed as the theory's
greatest merit. This is because the fast electron scales are not part of the
problem anymore: The plasma frequency and Debye length are not finite due to
charge-neutrality and neither is the electron Larmor motion due to the
vanishing electron mass. The absence of these fast time scales simplifies the
analysis and significantly reduces the computational burden in numerical
simulations \citep{Byers1978,Winske2003,Kunz2014}.

The electromagnetic field is described by the pre-Maxwell equations. Ampère's
law $\mu_0\vec{J}=\nabla\times\vec{B}$ thus lacks the displacement current.
The system of equations is closed with the definition of the current density
\begin{equation}
  \label{eq:current}
  \vec{J} = -en_e\vec{u}_e + \sums \qs\int d^3v\,\vec{v} f_s
\end{equation}
and the charge-neutrality condition
\begin{equation}
  \label{eq:charge-neutrality}
   en_e = \sums \qs\int d^3v\,f_s,
\end{equation}
where the summation is over ion species only.

The electron velocity $\ue$ plays a more fundamental role in Vlasov-fluid
theory than may be apparent at first sight. Differentiating
\cref{eq:charge-neutrality} with respect to time yields after a little bit of
algebra the continuity equation
\begin{equation}
  \label{eq:charge-continuity}
  \pder{n_e}{t} + \nabla\cdot(n_e\vec{u}_e) = 0.
\end{equation}
With the barotropic assumption, eliminating the electric field between Ohm's
law and Faraday's law yields the induction equation
\begin{equation}
  \label{eq:induction}
  \pder{\vec{B}}{t} = \nabla\times(\vec{u}_e\times\vec{B}).
\end{equation}
The magnetic field is thus frozen into the electron fluid. The form of
\cref{eq:charge-continuity,eq:induction} suggests that in linear theory we
introduce the Lagrangian displacement through
\begin{equation}
  \label{eq:delta-ue}
  \delta\ue + \vec{\xi}\cdot\nabla\ue =
  \left(\pder{}{t} + \ue\cdot\nabla\right)\vec{\xi}
\end{equation}
cf.\ \citet{Lynden-Bell1967,Friedman1978}. The linearized
\cref{eq:charge-continuity,eq:induction} are then easily integrated to yield
\begin{equation}
  \label{eq:delta-ne}
  \delta n_e + \nabla\cdot(n_e\vec{\xi}) = 0
\end{equation}
and
\begin{equation}
  \label{eq:delta-B}
  \delta\vec{B} = \nabla\times(\vec{\xi}\times\vec{B}).
\end{equation}
We stress that \cref{eq:delta-ue,eq:delta-ne,eq:delta-B} hold for general
equilibria.

\subsection{The dispersion relation in the shearing sheet}

In this section we linearize the Vlasov-fluid equations around the equilibrium
discussed in \cref{sec:steady-state}. The equilibrium bulk velocity common to
all species is thus given by the linear shear flow \labelcref{eq:shear-flow}.
As in \cref{sec:conductivity-tensor} we will work in terms of the invariant
velocity $\tvec{v}$ and electric field $\tvec{E}$ as defined in
\cref{eq:relative-quantities}. We thus use the Vlasov equation and Faraday's
law in the form of \cref{eq:vlasov-relative,eq:faraday-relative},
respectively. Also introducing the invariant electron velocity
\begin{equation}
  \tue = \ue + q\Omega x\ey,
\end{equation}
the generalized Ohm's law reads
\begin{equation}
  \label{eq:ohms-law-relative}
  e n_e(\tvec{E} + \tue\times\vec{B}) + \nabla\pe = 0,
\end{equation}
and the electric current density is given by
\begin{equation}
  \label{eq:ampere-relative}
  \vec{J} = -en_e\tilde{\vec{u}}_e + \sums \qs\int d^3v\,\tvec{v} f_s.
\end{equation}

The dispersion relation is obtained by linearizing \cref{eq:ampere-relative}
and expressing it entirely in terms of the Lagrangian displacement
$\vec{\xi}$. As in \cref{sec:conductivity-tensor} we consider an axisymmetric
plane wave. \Cref{eq:delta-ue} then reduces to
\begin{equation}
  \label{eq:delta-ue-sheet}
  \delta\tvec{u}_e = -i\omega\vec{\xi} + q\Omega\xi_x\ey.
\end{equation}
The ion current is given by the right hand side of
\cref{eq:conductive-relation}, to wit
\begin{equation}
  \sums \qs\int d^3v\,\tvec{v}\,\delta f_s
  = \left(
    \vec{\sigma} - \frac{1}{i\omega}\sums\frac{\qs^2 n_s}{m_s}\Delta_s\ey\ey
  \right)\cdot\left(\mathbf{1} - \frac{q\Omega}{i\omega}\ex\ey\right)
  \cdot\delta\tvec{E},
\end{equation}
where the ion conductivity tensor $\vec{\sigma}$ is defined in
\cref{eq:conductivity} with the understanding that the summation in this case
is over ion species only. The electric field is obtained from the generalized
Ohm's law in the form of \cref{eq:ohms-law-relative}. Using
\cref{eq:delta-ue-sheet,eq:delta-B} to eliminate the electron density and
velocity yields
\begin{equation}
  \label{eq:ohm-linear-sheet}
  \delta\tvec{E}
  = (i\omega\mathbf{1} + q\Omega\ex\ey)\cdot(\vec{\xi}\times\vec{B})
  + e^{-1}d\pe/dn_e\nabla\nabla\cdot\vec{\xi}.
\end{equation}
Finally, we combine Ampère's law and \cref{eq:delta-B} to obtain
\begin{equation}
  \mu_0\delta\vec{J} = \nabla\times[\nabla\times(\vec{\xi}\times\vec{B})].
\end{equation}
Putting everything together yields after some algebra
\begin{equation}
  \label{eq:ampere-xi}
  \mu_0^{-1}\nabla\times\nabla\times(\vec{\xi}\times\vec{B}) =
  en_e i\omega\vec{\xi} + \vec{\sigma}\cdot(i\omega\vec{\xi}\times\vec{B}
  + e^{-1}d\pe/dn_e\nabla\nabla\cdot\vec{\xi})
  - 2q\Omega^2\xi_x\ey\sums\frac{\qs n_s}{\omega_{cs}b_z + 2\Omega}.
\end{equation}
Assuming a plane wave decomposition, this equation may be written as
$\mathbf{D}\cdot\vec{\xi}=0$ with the dispersion tensor given by
\begin{equation}
  \label{eq:vlasov-fluid-dispersion-tensor}
  \mathbf{D} =
  \left[
    (k^2\mathbf{1} - \vec{k}\vec{k} - i\omega\mu_0\vec{\sigma})\va^2
    - i\omega\ome\vec{b}\times\mathbf{1}
    - 2q\Omega^2\sums\frac{n_s m_s}{\rho}
    \frac{\omcs/b_z}{\omcs b_z + 2\Omega}\ey\ey
  \right]\times\vec{b}
  + \ome\left(
    i\omega\vec{b}\vec{b}
    - \frac{\vec{\sigma}\cdot\vec{k}\vec{k}}{e^2n_e}\der{\pe}{n_e}
  \right),
\end{equation}
where
\begin{equation}
  \ome = \frac{en_e B}{\rho} = \sums\frac{n_s m_s}{\rho}\,\omcs
\end{equation}
is the mass-weighted mean cyclotron frequency. The perpendicular components of
\cref{eq:vlasov-fluid-dispersion-tensor} agree with
\cref{eq:perpendicular-dispersion}.

\bibliography{references}

\begin{thebibliography}{45}%
\makeatletter
\providecommand \@ifxundefined [1]{%
 \@ifx{#1\undefined}
}%
\providecommand \@ifnum [1]{%
 \ifnum #1\expandafter \@firstoftwo
 \else \expandafter \@secondoftwo
 \fi
}%
\providecommand \@ifx [1]{%
 \ifx #1\expandafter \@firstoftwo
 \else \expandafter \@secondoftwo
 \fi
}%
\providecommand \natexlab [1]{#1}%
\providecommand \enquote  [1]{``#1''}%
\providecommand \bibnamefont  [1]{#1}%
\providecommand \bibfnamefont [1]{#1}%
\providecommand \citenamefont [1]{#1}%
\providecommand \href@noop [0]{\@secondoftwo}%
\providecommand \href [0]{\begingroup \@sanitize@url \@href}%
\providecommand \@href[1]{\@@startlink{#1}\@@href}%
\providecommand \@@href[1]{\endgroup#1\@@endlink}%
\providecommand \@sanitize@url [0]{\catcode `\\12\catcode `\$12\catcode
  `\&12\catcode `\#12\catcode `\^12\catcode `\_12\catcode `\%12\relax}%
\providecommand \@@startlink[1]{}%
\providecommand \@@endlink[0]{}%
\providecommand \url  [0]{\begingroup\@sanitize@url \@url }%
\providecommand \@url [1]{\endgroup\@href {#1}{\urlprefix }}%
\providecommand \urlprefix  [0]{URL }%
\providecommand \Eprint [0]{\href }%
\providecommand \doibase [0]{http://dx.doi.org/}%
\providecommand \selectlanguage [0]{\@gobble}%
\providecommand \bibinfo  [0]{\@secondoftwo}%
\providecommand \bibfield  [0]{\@secondoftwo}%
\providecommand \translation [1]{[#1]}%
\providecommand \BibitemOpen [0]{}%
\providecommand \bibitemStop [0]{}%
\providecommand \bibitemNoStop [0]{.\EOS\space}%
\providecommand \EOS [0]{\spacefactor3000\relax}%
\providecommand \BibitemShut  [1]{\csname bibitem#1\endcsname}%
\let\auto@bib@innerbib\@empty
\bibitem [{\citenamefont {Hill}(1878)}]{Hill1878}%
  \BibitemOpen
  \bibfield  {author} {\bibinfo {author} {\bibfnamefont {G.~W.}\ \bibnamefont
  {Hill}},\ }\href {http://www.jstor.org/stable/2369304} {\bibfield  {journal}
  {\bibinfo  {journal} {American Journal of Mathematics}\ }\textbf {\bibinfo
  {volume} {1}},\ \bibinfo {pages} {129} (\bibinfo {year} {1878})}\BibitemShut
  {NoStop}%
\bibitem [{\citenamefont {Goldreich}\ and\ \citenamefont
  {Lynden-Bell}(1965)}]{Goldreich1965}%
  \BibitemOpen
  \bibfield  {author} {\bibinfo {author} {\bibfnamefont {P.}~\bibnamefont
  {Goldreich}}\ and\ \bibinfo {author} {\bibfnamefont {D.}~\bibnamefont
  {Lynden-Bell}},\ }\href@noop {} {\bibfield  {journal} {\bibinfo  {journal}
  {Monthly Notices of the Royal Astronomical Society}\ }\textbf {\bibinfo
  {volume} {130}},\ \bibinfo {pages} {125} (\bibinfo {year}
  {1965})}\BibitemShut {NoStop}%
\bibitem [{\citenamefont {Hawley}\ \emph {et~al.}(1996)\citenamefont {Hawley},
  \citenamefont {Gammie},\ and\ \citenamefont {Balbus}}]{Hawley1996}%
  \BibitemOpen
  \bibfield  {author} {\bibinfo {author} {\bibfnamefont {J.~F.}\ \bibnamefont
  {Hawley}}, \bibinfo {author} {\bibfnamefont {C.~F.}\ \bibnamefont {Gammie}},
  \ and\ \bibinfo {author} {\bibfnamefont {S.~A.}\ \bibnamefont {Balbus}},\
  }\href {http://adsabs.harvard.edu/doi/10.1086/177356} {\bibfield  {journal}
  {\bibinfo  {journal} {The Astrophysical Journal}\ }\textbf {\bibinfo {volume}
  {464}},\ \bibinfo {pages} {690} (\bibinfo {year} {1996})}\BibitemShut
  {NoStop}%
\bibitem [{\citenamefont {Rees}\ \emph {et~al.}(1982)\citenamefont {Rees},
  \citenamefont {Begelman}, \citenamefont {Blandford},\ and\ \citenamefont
  {Phinney}}]{Rees1982}%
  \BibitemOpen
  \bibfield  {author} {\bibinfo {author} {\bibfnamefont {M.~J.}\ \bibnamefont
  {Rees}}, \bibinfo {author} {\bibfnamefont {M.~C.}\ \bibnamefont {Begelman}},
  \bibinfo {author} {\bibfnamefont {R.~D.}\ \bibnamefont {Blandford}}, \ and\
  \bibinfo {author} {\bibfnamefont {E.~S.}\ \bibnamefont {Phinney}},\ }\href
  {http://www.nature.com/doifinder/10.1038/295017a0} {\bibfield  {journal}
  {\bibinfo  {journal} {Nature}\ }\textbf {\bibinfo {volume} {295}},\ \bibinfo
  {pages} {17} (\bibinfo {year} {1982})}\BibitemShut {NoStop}%
\bibitem [{\citenamefont {Yuan}\ and\ \citenamefont
  {Narayan}(2012)}]{Yuan2014}%
  \BibitemOpen
  \bibfield  {author} {\bibinfo {author} {\bibfnamefont {F.}~\bibnamefont
  {Yuan}}\ and\ \bibinfo {author} {\bibfnamefont {R.}~\bibnamefont {Narayan}},\
  }\href
  {http://www.annualreviews.org/doi/abs/10.1146/annurev-astro-082812-141003}
  {\bibfield  {journal} {\bibinfo  {journal} {Annual Review of Astronomy and
  Astrophysics}\ }\textbf {\bibinfo {volume} {52}} (\bibinfo {year}
  {2012})}\BibitemShut {NoStop}%
\bibitem [{\citenamefont {Julian}\ and\ \citenamefont
  {Toomre}(1966)}]{Julian1966}%
  \BibitemOpen
  \bibfield  {author} {\bibinfo {author} {\bibfnamefont {W.~H.}\ \bibnamefont
  {Julian}}\ and\ \bibinfo {author} {\bibfnamefont {A.}~\bibnamefont
  {Toomre}},\ }\href {http://adsabs.harvard.edu/doi/10.1086/148957} {\bibfield
  {journal} {\bibinfo  {journal} {The Astrophysical Journal}\ }\textbf
  {\bibinfo {volume} {146}},\ \bibinfo {pages} {810} (\bibinfo {year}
  {1966})}\BibitemShut {NoStop}%
\bibitem [{\citenamefont {Ichimaru}(1973)}]{Ichimaru1973}%
  \BibitemOpen
  \bibfield  {author} {\bibinfo {author} {\bibfnamefont {S.}~\bibnamefont
  {Ichimaru}},\ }\href@noop {} {\emph {\bibinfo {title} {{Basic principles of
  plasma physics: a statistical approach}}}}\ (\bibinfo  {publisher}
  {Benjamin/Cummings Publishing Company},\ \bibinfo {year} {1973})\BibitemShut
  {NoStop}%
\bibitem [{\citenamefont {Krall}\ and\ \citenamefont
  {Trivelpiece}(1973)}]{Krall1973}%
  \BibitemOpen
  \bibfield  {author} {\bibinfo {author} {\bibfnamefont {N.~A.}\ \bibnamefont
  {Krall}}\ and\ \bibinfo {author} {\bibfnamefont {A.~W.}\ \bibnamefont
  {Trivelpiece}},\ }\href@noop {} {\emph {\bibinfo {title} {{Principles of
  plasma physics}}}}\ (\bibinfo  {publisher} {McGraw-Hill},\ \bibinfo {address}
  {New York},\ \bibinfo {year} {1973})\BibitemShut {NoStop}%
\bibitem [{\citenamefont {Stix}(1992)}]{Stix1992}%
  \BibitemOpen
  \bibfield  {author} {\bibinfo {author} {\bibfnamefont {T.~H.}\ \bibnamefont
  {Stix}},\ }\href@noop {} {\emph {\bibinfo {title} {{Waves in Plasmas}}}}\
  (\bibinfo  {publisher} {American Institute of Physics},\ \bibinfo {address}
  {New York},\ \bibinfo {year} {1992})\BibitemShut {NoStop}%
\bibitem [{\citenamefont {Quataert}\ \emph {et~al.}(2002)\citenamefont
  {Quataert}, \citenamefont {Dorland},\ and\ \citenamefont
  {Hammett}}]{Quataert2002}%
  \BibitemOpen
  \bibfield  {author} {\bibinfo {author} {\bibfnamefont {E.}~\bibnamefont
  {Quataert}}, \bibinfo {author} {\bibfnamefont {W.}~\bibnamefont {Dorland}}, \
  and\ \bibinfo {author} {\bibfnamefont {G.~W.}\ \bibnamefont {Hammett}},\
  }\href {http://stacks.iop.org/0004-637X/577/i=1/a=524} {\bibfield  {journal}
  {\bibinfo  {journal} {The Astrophysical Journal}\ }\textbf {\bibinfo {volume}
  {577}},\ \bibinfo {pages} {524} (\bibinfo {year} {2002})}\BibitemShut
  {NoStop}%
\bibitem [{\citenamefont {Wardle}(1999)}]{Wardle1999}%
  \BibitemOpen
  \bibfield  {author} {\bibinfo {author} {\bibfnamefont {M.}~\bibnamefont
  {Wardle}},\ }\href
  {http://mnras.oxfordjournals.org/cgi/doi/10.1046/j.1365-8711.1999.02670.x}
  {\bibfield  {journal} {\bibinfo  {journal} {Monthly Notices of the Royal
  Astronomical Society}\ }\textbf {\bibinfo {volume} {307}},\ \bibinfo {pages}
  {849} (\bibinfo {year} {1999})}\BibitemShut {NoStop}%
\bibitem [{\citenamefont {Ferraro}(2007)}]{Ferraro2007}%
  \BibitemOpen
  \bibfield  {author} {\bibinfo {author} {\bibfnamefont {N.~M.}\ \bibnamefont
  {Ferraro}},\ }\href {http://stacks.iop.org/0004-637X/662/i=1/a=512}
  {\bibfield  {journal} {\bibinfo  {journal} {The Astrophysical Journal}\
  }\textbf {\bibinfo {volume} {662}},\ \bibinfo {pages} {512} (\bibinfo {year}
  {2007})}\BibitemShut {NoStop}%
\bibitem [{\citenamefont {Krolik}\ and\ \citenamefont
  {Zweibel}(2006)}]{Krolik2006}%
  \BibitemOpen
  \bibfield  {author} {\bibinfo {author} {\bibfnamefont {J.~H.}\ \bibnamefont
  {Krolik}}\ and\ \bibinfo {author} {\bibfnamefont {E.~G.}\ \bibnamefont
  {Zweibel}},\ }\href {http://stacks.iop.org/0004-637X/644/i=1/a=651}
  {\bibfield  {journal} {\bibinfo  {journal} {The Astrophysical Journal}\
  }\textbf {\bibinfo {volume} {644}},\ \bibinfo {pages} {651} (\bibinfo {year}
  {2006})}\BibitemShut {NoStop}%
\bibitem [{\citenamefont {Wisdom}\ and\ \citenamefont
  {Tremaine}(1988)}]{Wisdom1988}%
  \BibitemOpen
  \bibfield  {author} {\bibinfo {author} {\bibfnamefont {J.}~\bibnamefont
  {Wisdom}}\ and\ \bibinfo {author} {\bibfnamefont {S.}~\bibnamefont
  {Tremaine}},\ }\href
  {http://adsabs.harvard.edu/cgi-bin/bib\_query?1988AJ.....95..925W} {\bibfield
   {journal} {\bibinfo  {journal} {The Astronomical Journal}\ }\textbf
  {\bibinfo {volume} {95}},\ \bibinfo {pages} {925} (\bibinfo {year}
  {1988})}\BibitemShut {NoStop}%
\bibitem [{\citenamefont {Ferraro}(1937)}]{Ferraro1937}%
  \BibitemOpen
  \bibfield  {author} {\bibinfo {author} {\bibfnamefont {V.~C.~A.}\
  \bibnamefont {Ferraro}},\ }\href@noop {} {\bibfield  {journal} {\bibinfo
  {journal} {Monthly Notices of the Royal Astronomical Society}\ }\textbf
  {\bibinfo {volume} {97}},\ \bibinfo {pages} {458} (\bibinfo {year}
  {1937})}\BibitemShut {NoStop}%
\bibitem [{\citenamefont {Shu}(1969)}]{Shu1969}%
  \BibitemOpen
  \bibfield  {author} {\bibinfo {author} {\bibfnamefont {F.~H.}\ \bibnamefont
  {Shu}},\ }\href {http://adsabs.harvard.edu/doi/10.1086/150214} {\bibfield
  {journal} {\bibinfo  {journal} {The Astrophysical Journal}\ }\textbf
  {\bibinfo {volume} {158}},\ \bibinfo {pages} {505} (\bibinfo {year}
  {1969})}\BibitemShut {NoStop}%
\bibitem [{\citenamefont {Kaufman}(1971)}]{Kaufman1971}%
  \BibitemOpen
  \bibfield  {author} {\bibinfo {author} {\bibfnamefont {A.~N.}\ \bibnamefont
  {Kaufman}},\ }\href {http://link.aip.org/link/PFLDAS/v14/i2/p387/s1\&Agg=doi}
  {\bibfield  {journal} {\bibinfo  {journal} {Physics of Fluids}\ }\textbf
  {\bibinfo {volume} {14}},\ \bibinfo {pages} {387} (\bibinfo {year}
  {1971})}\BibitemShut {NoStop}%
\bibitem [{\citenamefont {Kaufman}(1972)}]{Kaufman1972}%
  \BibitemOpen
  \bibfield  {author} {\bibinfo {author} {\bibfnamefont {A.~N.}\ \bibnamefont
  {Kaufman}},\ }\href
  {http://link.aip.org/link/PFLDAS/v15/i6/p1063/s1\&Agg=doi} {\bibfield
  {journal} {\bibinfo  {journal} {Physics of Fluids}\ }\textbf {\bibinfo
  {volume} {15}},\ \bibinfo {pages} {1063} (\bibinfo {year}
  {1972})}\BibitemShut {NoStop}%
\bibitem [{\citenamefont {Thomson}(1887)}]{Thomson1887}%
  \BibitemOpen
  \bibfield  {author} {\bibinfo {author} {\bibfnamefont {W.}~\bibnamefont
  {Thomson}},\ }\href
  {http://www.tandfonline.com/doi/abs/10.1080/14786448708628078} {\bibfield
  {journal} {\bibinfo  {journal} {Philosophical Magazine Series 5}\ }\textbf
  {\bibinfo {volume} {24}},\ \bibinfo {pages} {188} (\bibinfo {year}
  {1887})}\BibitemShut {NoStop}%
\bibitem [{\citenamefont {Lees}\ and\ \citenamefont
  {Edwards}(1972)}]{Lees1972}%
  \BibitemOpen
  \bibfield  {author} {\bibinfo {author} {\bibfnamefont {A.~W.}\ \bibnamefont
  {Lees}}\ and\ \bibinfo {author} {\bibfnamefont {S.~F.}\ \bibnamefont
  {Edwards}},\ }\href
  {http://stacks.iop.org/0022-3719/5/i=15/a=006?key=crossref.e594e5cd0b6eb0d5dc9f68b23ee836be}
  {\bibfield  {journal} {\bibinfo  {journal} {Journal of Physics C: Solid State
  Physics}\ }\textbf {\bibinfo {volume} {5}},\ \bibinfo {pages} {1921}
  (\bibinfo {year} {1972})}\BibitemShut {NoStop}%
\bibitem [{\citenamefont {Hawley}\ \emph {et~al.}(1995)\citenamefont {Hawley},
  \citenamefont {Gammie},\ and\ \citenamefont {Balbus}}]{Hawley1995}%
  \BibitemOpen
  \bibfield  {author} {\bibinfo {author} {\bibfnamefont {J.~F.}\ \bibnamefont
  {Hawley}}, \bibinfo {author} {\bibfnamefont {C.~F.}\ \bibnamefont {Gammie}},
  \ and\ \bibinfo {author} {\bibfnamefont {S.~A.}\ \bibnamefont {Balbus}},\
  }\href {http://adsabs.harvard.edu/doi/10.1086/175311} {\bibfield  {journal}
  {\bibinfo  {journal} {The Astrophysical Journal}\ }\textbf {\bibinfo {volume}
  {440}},\ \bibinfo {pages} {742} (\bibinfo {year} {1995})}\BibitemShut
  {NoStop}%
\bibitem [{\citenamefont {Fried}\ and\ \citenamefont
  {Conte}(1961)}]{Fried1961}%
  \BibitemOpen
  \bibfield  {author} {\bibinfo {author} {\bibfnamefont {B.~D.}\ \bibnamefont
  {Fried}}\ and\ \bibinfo {author} {\bibfnamefont {S.~D.}\ \bibnamefont
  {Conte}},\ }\href@noop {} {\emph {\bibinfo {title} {{The Plasma Dispersion
  Function}}}}\ (\bibinfo  {publisher} {Academic Press},\ \bibinfo {address}
  {New York},\ \bibinfo {year} {1961})\BibitemShut {NoStop}%
\bibitem [{\citenamefont {Balbus}\ and\ \citenamefont
  {Terquem}(2001)}]{Balbus2001}%
  \BibitemOpen
  \bibfield  {author} {\bibinfo {author} {\bibfnamefont {S.~A.}\ \bibnamefont
  {Balbus}}\ and\ \bibinfo {author} {\bibfnamefont {C.}~\bibnamefont
  {Terquem}},\ }\href {http://stacks.iop.org/0004-637X/552/i=1/a=235}
  {\bibfield  {journal} {\bibinfo  {journal} {The Astrophysical Journal}\
  }\textbf {\bibinfo {volume} {552}},\ \bibinfo {pages} {235} (\bibinfo {year}
  {2001})}\BibitemShut {NoStop}%
\bibitem [{\citenamefont {Grad}(1961)}]{Grad1961}%
  \BibitemOpen
  \bibfield  {author} {\bibinfo {author} {\bibfnamefont {H.}~\bibnamefont
  {Grad}},\ }\href@noop {} {\emph {\bibinfo {title} {{Microscopic and
  Macroscopic Models in Plasma Physics}}}},\ \bibinfo {type} {Tech. Rep.}\
  (\bibinfo {year} {1961})\BibitemShut {NoStop}%
\bibitem [{\citenamefont {Freidberg}(1972)}]{Freidberg1972}%
  \BibitemOpen
  \bibfield  {author} {\bibinfo {author} {\bibfnamefont {J.~P.}\ \bibnamefont
  {Freidberg}},\ }\href
  {http://link.aip.org/link/PFLDAS/v15/i6/p1102/s1\&Agg=doi} {\bibfield
  {journal} {\bibinfo  {journal} {Physics of Fluids}\ }\textbf {\bibinfo
  {volume} {15}},\ \bibinfo {pages} {1102} (\bibinfo {year}
  {1972})}\BibitemShut {NoStop}%
\bibitem [{\citenamefont {Kulsrud}(1983)}]{Kulsrud1983}%
  \BibitemOpen
  \bibfield  {author} {\bibinfo {author} {\bibfnamefont {R.~M.}\ \bibnamefont
  {Kulsrud}},\ }in\ \href@noop {} {\emph {\bibinfo {booktitle} {Handbook of
  Plasma Physics}}},\ Vol.~\bibinfo {volume} {1},\ \bibinfo {editor} {edited
  by\ \bibinfo {editor} {\bibfnamefont {M.~N.}\ \bibnamefont {Rosenbluth}}\
  and\ \bibinfo {editor} {\bibfnamefont {R.}~\bibnamefont {Sagdeev}}}\
  (\bibinfo  {publisher} {North Holland},\ \bibinfo {address} {New York},\
  \bibinfo {year} {1983})\ Chap.\ \bibinfo {chapter} {1.4}, pp.\ \bibinfo
  {pages} {115--145}\BibitemShut {NoStop}%
\bibitem [{\citenamefont {Hazeltine}\ and\ \citenamefont
  {Waelbroeck}(2004)}]{Hazeltine2004}%
  \BibitemOpen
  \bibfield  {author} {\bibinfo {author} {\bibfnamefont {R.~D.}\ \bibnamefont
  {Hazeltine}}\ and\ \bibinfo {author} {\bibfnamefont {F.~L.}\ \bibnamefont
  {Waelbroeck}},\ }\href@noop {} {\emph {\bibinfo {title} {{The Framework of
  Plasma Physics}}}}\ (\bibinfo  {publisher} {Westview},\ \bibinfo {address}
  {Boulder, CO},\ \bibinfo {year} {2004})\BibitemShut {NoStop}%
\bibitem [{\citenamefont {Balbus}(2004)}]{Balbus2004}%
  \BibitemOpen
  \bibfield  {author} {\bibinfo {author} {\bibfnamefont {S.~A.}\ \bibnamefont
  {Balbus}},\ }\href {http://stacks.iop.org/0004-637X/616/i=2/a=857} {\bibfield
   {journal} {\bibinfo  {journal} {The Astrophysical Journal}\ }\textbf
  {\bibinfo {volume} {616}},\ \bibinfo {pages} {857} (\bibinfo {year}
  {2004})}\BibitemShut {NoStop}%
\bibitem [{\citenamefont {Schekochihin}\ \emph {et~al.}(2005)\citenamefont
  {Schekochihin}, \citenamefont {Cowley}, \citenamefont {Kulsrud},
  \citenamefont {Hammett},\ and\ \citenamefont {Sharma}}]{Schekochihin2005}%
  \BibitemOpen
  \bibfield  {author} {\bibinfo {author} {\bibfnamefont {A.~A.}\ \bibnamefont
  {Schekochihin}}, \bibinfo {author} {\bibfnamefont {S.~C.}\ \bibnamefont
  {Cowley}}, \bibinfo {author} {\bibfnamefont {R.~M.}\ \bibnamefont {Kulsrud}},
  \bibinfo {author} {\bibfnamefont {G.~W.}\ \bibnamefont {Hammett}}, \ and\
  \bibinfo {author} {\bibfnamefont {P.}~\bibnamefont {Sharma}},\ }\href
  {http://stacks.iop.org/0004-637X/629/i=1/a=139} {\bibfield  {journal}
  {\bibinfo  {journal} {The Astrophysical Journal}\ }\textbf {\bibinfo {volume}
  {629}},\ \bibinfo {pages} {139} (\bibinfo {year} {2005})}\BibitemShut
  {NoStop}%
\bibitem [{\citenamefont {Sharma}\ \emph {et~al.}(2006)\citenamefont {Sharma},
  \citenamefont {Hammett}, \citenamefont {Quataert},\ and\ \citenamefont
  {Stone}}]{Sharma2006}%
  \BibitemOpen
  \bibfield  {author} {\bibinfo {author} {\bibfnamefont {P.}~\bibnamefont
  {Sharma}}, \bibinfo {author} {\bibfnamefont {G.~W.}\ \bibnamefont {Hammett}},
  \bibinfo {author} {\bibfnamefont {E.}~\bibnamefont {Quataert}}, \ and\
  \bibinfo {author} {\bibfnamefont {J.~M.}\ \bibnamefont {Stone}},\ }\href
  {http://stacks.iop.org/0004-637X/637/i=2/a=952} {\bibfield  {journal}
  {\bibinfo  {journal} {The Astrophysical Journal}\ }\textbf {\bibinfo {volume}
  {637}},\ \bibinfo {pages} {952} (\bibinfo {year} {2006})}\BibitemShut
  {NoStop}%
\bibitem [{\citenamefont {Grad}(1966)}]{Grad1966}%
  \BibitemOpen
  \bibfield  {author} {\bibinfo {author} {\bibfnamefont {H.}~\bibnamefont
  {Grad}},\ }\href {http://www.archive.org/details/guidingcenterpla00grad}
  {\emph {\bibinfo {title} {{The Guiding Center Plasma}}}},\ \bibinfo {type}
  {Tech. Rep.}\ (\bibinfo {year} {1966})\BibitemShut {NoStop}%
\bibitem [{\citenamefont {Collins}\ \emph {et~al.}(2012)\citenamefont
  {Collins}, \citenamefont {Katz}, \citenamefont {Wallace}, \citenamefont
  {Jara-Almonte}, \citenamefont {Reese}, \citenamefont {Zweibel},\ and\
  \citenamefont {Forest}}]{Collins2012}%
  \BibitemOpen
  \bibfield  {author} {\bibinfo {author} {\bibfnamefont {C.}~\bibnamefont
  {Collins}}, \bibinfo {author} {\bibfnamefont {N.}~\bibnamefont {Katz}},
  \bibinfo {author} {\bibfnamefont {J.}~\bibnamefont {Wallace}}, \bibinfo
  {author} {\bibfnamefont {J.}~\bibnamefont {Jara-Almonte}}, \bibinfo {author}
  {\bibfnamefont {I.}~\bibnamefont {Reese}}, \bibinfo {author} {\bibfnamefont
  {E.}~\bibnamefont {Zweibel}}, \ and\ \bibinfo {author} {\bibfnamefont
  {C.~B.}\ \bibnamefont {Forest}},\ }\href
  {http://link.aps.org/doi/10.1103/PhysRevLett.108.115001} {\bibfield
  {journal} {\bibinfo  {journal} {Physical Review Letters}\ }\textbf {\bibinfo
  {volume} {108}},\ \bibinfo {pages} {115001} (\bibinfo {year}
  {2012})}\BibitemShut {NoStop}%
\bibitem [{\citenamefont {Cary}\ and\ \citenamefont
  {Littlejohn}(1983)}]{Cary1983}%
  \BibitemOpen
  \bibfield  {author} {\bibinfo {author} {\bibfnamefont {J.~R.}\ \bibnamefont
  {Cary}}\ and\ \bibinfo {author} {\bibfnamefont {R.~G.}\ \bibnamefont
  {Littlejohn}},\ }\href
  {http://linkinghub.elsevier.com/retrieve/pii/0003491683903135} {\bibfield
  {journal} {\bibinfo  {journal} {Annals of Physics}\ }\textbf {\bibinfo
  {volume} {151}},\ \bibinfo {pages} {1} (\bibinfo {year} {1983})}\BibitemShut
  {NoStop}%
\bibitem [{\citenamefont {Papaloizou}\ and\ \citenamefont
  {Szuszkiewicz}(1992)}]{Papaloizou1992}%
  \BibitemOpen
  \bibfield  {author} {\bibinfo {author} {\bibfnamefont {J.~C.~B.}\
  \bibnamefont {Papaloizou}}\ and\ \bibinfo {author} {\bibfnamefont
  {E.}~\bibnamefont {Szuszkiewicz}},\ }\href
  {http://www.tandfonline.com/doi/abs/10.1080/03091929208229058} {\bibfield
  {journal} {\bibinfo  {journal} {Geophysical \& Astrophysical Fluid Dynamics}\
  }\textbf {\bibinfo {volume} {66}},\ \bibinfo {pages} {223} (\bibinfo {year}
  {1992})}\BibitemShut {NoStop}%
\bibitem [{\citenamefont {Krechetnikov}\ and\ \citenamefont
  {Marsden}(2007)}]{Krechetnikov2007}%
  \BibitemOpen
  \bibfield  {author} {\bibinfo {author} {\bibfnamefont {R.}~\bibnamefont
  {Krechetnikov}}\ and\ \bibinfo {author} {\bibfnamefont {J.~E.}\ \bibnamefont
  {Marsden}},\ }\href {http://link.aps.org/doi/10.1103/RevModPhys.79.519}
  {\bibfield  {journal} {\bibinfo  {journal} {Reviews of Modern Physics}\
  }\textbf {\bibinfo {volume} {79}},\ \bibinfo {pages} {519} (\bibinfo {year}
  {2007})}\BibitemShut {NoStop}%
\bibitem [{\citenamefont {Chandrasekhar}(1961)}]{Chandrasekhar1961}%
  \BibitemOpen
  \bibfield  {author} {\bibinfo {author} {\bibfnamefont {S.}~\bibnamefont
  {Chandrasekhar}},\ }\href@noop {} {\emph {\bibinfo {title} {International
  Series of Monographs on Physics, Oxford: Clarendon, 1961}}}\ (\bibinfo
  {publisher} {Dover Publications},\ \bibinfo {address} {Oxford},\ \bibinfo
  {year} {1961})\BibitemShut {NoStop}%
\bibitem [{\citenamefont {Holm}\ \emph {et~al.}(1985)\citenamefont {Holm},
  \citenamefont {Marsden}, \citenamefont {Ratiu},\ and\ \citenamefont
  {Weinstein}}]{Holm1985}%
  \BibitemOpen
  \bibfield  {author} {\bibinfo {author} {\bibfnamefont {D.~D.}\ \bibnamefont
  {Holm}}, \bibinfo {author} {\bibfnamefont {J.~E.}\ \bibnamefont {Marsden}},
  \bibinfo {author} {\bibfnamefont {T.}~\bibnamefont {Ratiu}}, \ and\ \bibinfo
  {author} {\bibfnamefont {A.}~\bibnamefont {Weinstein}},\ }\href
  {http://linkinghub.elsevier.com/retrieve/pii/0370157385900286} {\bibfield
  {journal} {\bibinfo  {journal} {Physics Reports}\ }\textbf {\bibinfo {volume}
  {123}},\ \bibinfo {pages} {1} (\bibinfo {year} {1985})}\BibitemShut {NoStop}%
\bibitem [{\citenamefont {Bloch}\ \emph {et~al.}(1994)\citenamefont {Bloch},
  \citenamefont {Krishnaprasad}, \citenamefont {Marsden},\ and\ \citenamefont
  {Ratiu}}]{Bloch1994}%
  \BibitemOpen
  \bibfield  {author} {\bibinfo {author} {\bibfnamefont {A.}~\bibnamefont
  {Bloch}}, \bibinfo {author} {\bibfnamefont {P.~S.}\ \bibnamefont
  {Krishnaprasad}}, \bibinfo {author} {\bibfnamefont {J.~E.}\ \bibnamefont
  {Marsden}}, \ and\ \bibinfo {author} {\bibfnamefont {T.}~\bibnamefont
  {Ratiu}},\ }\href
  {http://resolver.caltech.edu/CaltechAUTHORS:20100907-102644978} {\bibfield
  {journal} {\bibinfo  {journal} {Annales de l'Institut Henri Poincar\'{e} (C)
  Analyse Non Lin\'{e}aire}\ }\textbf {\bibinfo {volume} {11}},\ \bibinfo
  {pages} {37} (\bibinfo {year} {1994})}\BibitemShut {NoStop}%
\bibitem [{\citenamefont {Cerfon}\ and\ \citenamefont
  {Freidberg}(2011)}]{Cerfon2011}%
  \BibitemOpen
  \bibfield  {author} {\bibinfo {author} {\bibfnamefont {A.~J.}\ \bibnamefont
  {Cerfon}}\ and\ \bibinfo {author} {\bibfnamefont {J.~P.}\ \bibnamefont
  {Freidberg}},\ }\href
  {http://link.aip.org/link/PHPAEN/v18/i1/p012505/s1\&Agg=doi} {\bibfield
  {journal} {\bibinfo  {journal} {Physics of Plasmas}\ }\textbf {\bibinfo
  {volume} {18}},\ \bibinfo {pages} {012505} (\bibinfo {year}
  {2011})}\BibitemShut {NoStop}%
\bibitem [{\citenamefont {Rosin}\ \emph {et~al.}(2011)\citenamefont {Rosin},
  \citenamefont {Schekochihin}, \citenamefont {Rincon},\ and\ \citenamefont
  {Cowley}}]{Rosin2011}%
  \BibitemOpen
  \bibfield  {author} {\bibinfo {author} {\bibfnamefont {M.~S.}\ \bibnamefont
  {Rosin}}, \bibinfo {author} {\bibfnamefont {A.~A.}\ \bibnamefont
  {Schekochihin}}, \bibinfo {author} {\bibfnamefont {F.}~\bibnamefont
  {Rincon}}, \ and\ \bibinfo {author} {\bibfnamefont {S.~C.}\ \bibnamefont
  {Cowley}},\ }\href {http://doi.wiley.com/10.1111/j.1365-2966.2010.17931.x}
  {\bibfield  {journal} {\bibinfo  {journal} {Monthly Notices of the Royal
  Astronomical Society}\ }\textbf {\bibinfo {volume} {413}},\ \bibinfo {pages}
  {7} (\bibinfo {year} {2011})}\BibitemShut {NoStop}%
\bibitem [{\citenamefont {Byers}\ \emph {et~al.}(1978)\citenamefont {Byers},
  \citenamefont {Cohen}, \citenamefont {Condit},\ and\ \citenamefont
  {Hanson}}]{Byers1978}%
  \BibitemOpen
  \bibfield  {author} {\bibinfo {author} {\bibfnamefont {J.~A.}\ \bibnamefont
  {Byers}}, \bibinfo {author} {\bibfnamefont {B.~I.}\ \bibnamefont {Cohen}},
  \bibinfo {author} {\bibfnamefont {W.~C.}\ \bibnamefont {Condit}}, \ and\
  \bibinfo {author} {\bibfnamefont {J.~D.}\ \bibnamefont {Hanson}},\ }\href
  {http://linkinghub.elsevier.com/retrieve/pii/0021999178900165} {\bibfield
  {journal} {\bibinfo  {journal} {Journal of Computational Physics}\ }\textbf
  {\bibinfo {volume} {27}},\ \bibinfo {pages} {363} (\bibinfo {year}
  {1978})}\BibitemShut {NoStop}%
\bibitem [{\citenamefont {Winske}\ \emph {et~al.}(2003)\citenamefont {Winske},
  \citenamefont {Yin}, \citenamefont {Omidi}, \citenamefont {Karimabadi},\ and\
  \citenamefont {Quest}}]{Winske2003}%
  \BibitemOpen
  \bibfield  {author} {\bibinfo {author} {\bibfnamefont {D.}~\bibnamefont
  {Winske}}, \bibinfo {author} {\bibfnamefont {L.}~\bibnamefont {Yin}},
  \bibinfo {author} {\bibfnamefont {N.}~\bibnamefont {Omidi}}, \bibinfo
  {author} {\bibfnamefont {H.}~\bibnamefont {Karimabadi}}, \ and\ \bibinfo
  {author} {\bibfnamefont {K.}~\bibnamefont {Quest}},\ }in\ \href
  {http://www.springerlink.com/content/30w1142154n11651/} {\emph {\bibinfo
  {booktitle} {Lecture Notes in Physics: Space Plasma Simulation}}},\ \bibinfo
  {editor} {edited by\ \bibinfo {editor} {\bibfnamefont {J.}~\bibnamefont
  {B\"{u}chner}}, \bibinfo {editor} {\bibfnamefont {M.}~\bibnamefont
  {Scholer}}, \ and\ \bibinfo {editor} {\bibfnamefont {C.}~\bibnamefont
  {Dum}}}\ (\bibinfo  {publisher} {Springer Berlin/Heidelberg},\ \bibinfo
  {year} {2003})\ pp.\ \bibinfo {pages} {136--165}\BibitemShut {NoStop}%
\bibitem [{\citenamefont {Kunz}\ \emph {et~al.}(2014)\citenamefont {Kunz},
  \citenamefont {Stone},\ and\ \citenamefont {Bai}}]{Kunz2014}%
  \BibitemOpen
  \bibfield  {author} {\bibinfo {author} {\bibfnamefont {M.~W.}\ \bibnamefont
  {Kunz}}, \bibinfo {author} {\bibfnamefont {J.~M.}\ \bibnamefont {Stone}}, \
  and\ \bibinfo {author} {\bibfnamefont {X.-N.}\ \bibnamefont {Bai}},\ }\href
  {http://linkinghub.elsevier.com/retrieve/pii/S0021999113007973} {\bibfield
  {journal} {\bibinfo  {journal} {Journal of Computational Physics}\ }\textbf
  {\bibinfo {volume} {259}},\ \bibinfo {pages} {154} (\bibinfo {year}
  {2014})}\BibitemShut {NoStop}%
\bibitem [{\citenamefont {Lynden-Bell}\ and\ \citenamefont
  {Ostriker}(1967)}]{Lynden-Bell1967}%
  \BibitemOpen
  \bibfield  {author} {\bibinfo {author} {\bibfnamefont {D.}~\bibnamefont
  {Lynden-Bell}}\ and\ \bibinfo {author} {\bibfnamefont {J.~P.}\ \bibnamefont
  {Ostriker}},\ }\href@noop {} {\bibfield  {journal} {\bibinfo  {journal}
  {Monthly Notices of the Royal Astronomical Society}\ }\textbf {\bibinfo
  {volume} {136}},\ \bibinfo {pages} {293} (\bibinfo {year}
  {1967})}\BibitemShut {NoStop}%
\bibitem [{\citenamefont {Friedman}\ and\ \citenamefont
  {Schutz}(1978)}]{Friedman1978}%
  \BibitemOpen
  \bibfield  {author} {\bibinfo {author} {\bibfnamefont {J.~L.}\ \bibnamefont
  {Friedman}}\ and\ \bibinfo {author} {\bibfnamefont {B.~F.}\ \bibnamefont
  {Schutz}},\ }\href {http://adsabs.harvard.edu/doi/10.1086/156098} {\bibfield
  {journal} {\bibinfo  {journal} {The Astrophysical Journal}\ }\textbf
  {\bibinfo {volume} {221}},\ \bibinfo {pages} {937} (\bibinfo {year}
  {1978})}\BibitemShut {NoStop}%
\end{thebibliography}%

\end{document}